\documentclass[letterpaper]{article} 
\usepackage[]{aaai23}  
\usepackage{times}  
\usepackage{helvet}  
\usepackage{courier}  
\usepackage[hyphens]{url}  
\usepackage{graphicx} 
\usepackage{natbib}  
\usepackage{caption} 
\usepackage{algorithm}
\usepackage{algorithmic}

\usepackage{verbatim}
\usepackage[]{appendix} 

\usepackage{newfloat}
\usepackage{listings}
\DeclareCaptionStyle{ruled}{labelfont=normalfont,labelsep=colon,strut=off} 
\floatstyle{ruled}
\newfloat{listing}{tb}{lst}{}
\floatname{listing}{Listing}
\nocopyright 

\usepackage{caption}
\usepackage{subcaption}
\usepackage{array}
\usepackage{makecell}
\usepackage{graphicx}

\title{A Friendly Face: Do Text-to-Image Systems Rely on Stereotypes when the Input is Under-Specified?}

\author{
Kathleen C. Fraser, Svetlana Kiritchenko, and Isar Nejadgholi
}
\affiliations{
   National Research Council Canada\\

    Ottawa, Canada\\
    \{kathleen.fraser, svetlana.kiritchenko, isar.nejadgholi\}@nrc-cnrc.gc.ca
}

\begin{document}

\maketitle

\begin{abstract}

As text-to-image systems continue to grow in popularity with the  general public, questions have arisen about bias and diversity in the generated images. Here, we investigate properties  of images generated in response to prompts which are visually under-specified, but contain salient social attributes (e.g., `a portrait of a threatening person' versus `a portrait of a friendly person'). Grounding our work in social cognition theory, we find that in many cases, images  contain similar  demographic biases to those reported in the stereotype literature. However, trends are inconsistent across different models and further investigation is warranted.

\end{abstract}

\section{Introduction}

Recent advances in natural language processing and computer vision have led to the development of text-to-image systems 
with unprecedented levels of realism and flexibility. 
At the same time, commentators have noted potential ethical issues related to the use of copyrighted artworks in the training sets, the generation of hateful and offensive content, as well as issues of bias and diversity in the model outputs. Relating to the latter, research work has begun to audit the output of such models, investigating stereotypical associations between occupations and particular races and genders \citep{cho2022dall}, as well as between the word ``American'' and lighter skin colours \citep{wolfe2022american}. 

Here, we take an alternative approach inspired by stereotype research in social psychology and focus on \textit{perceived traits} of individuals. However, we approach the problem from the inverse direction of most psychological studies. Rather than treating a demographic group (say, women) as the independent variable, and asking respondents for the associated traits (say, nurturing and emotional), here we use the \textit{trait} as a prompt to the text-to-image system, and observe the demographic properties of the resulting image. So, to continue our example: if we ask the system for an image of \textit{an emotional person}, will it return mostly pictures of women?

We ground our investigation in the ABC Model of social cognition \cite{koch2016abc}. This model proposes three basic dimensions of social judgement; namely: Agency (A), Beliefs (B), and Communion (C). These three dimensions can be further broken down into 16 polar traits. For example, Agency comprises traits such as \textit{powerful vs.\@ powerless} and \textit{high-status vs.\@ low-status}, Beliefs include traits such as \textit{conservative vs.\@ liberal} and \textit{religious vs.\@ science-oriented}, and Communion includes \textit{sincere vs.\@ dishonest} and \textit{altruistic vs.\@ egoistic}. This model
 suggests that all our stereotypes of different groups can be specified in this 3-dimensional space: e.g., in the North American context, Southerners may be stereotyped as laid-back, friendly, and religious (low agency, high communion, low beliefs\footnote{While Agency and Communion have clear positive (high) and negative (low) poles, the Beliefs dimension is defined along a continuum from \textit{progressive} to \textit{conservative}, with \textit{progressive} being arbitrarily assigned the ``positive'' direction. Note also that polarity does not necessarily align with normative judgements of \textit{good/bad} behaviour; e.g., dominating people have \textit{positive} agency, although their dominating behaviour would not necessarily be seen as \textit{good}.}), while tech entrepreneurs may be stereotyped as wealthy, science-oriented, and greedy (high agency, high beliefs, low communion). 

Clearly, these adjectives are under-specified with respect to a visual representation: what does a \textit{powerful} person look like? What does a \textit{sincere} person look like? It is precisely this under-specificity that can result in biased outputs, as the model must ``fill in the blanks'' with whatever cultural knowledge it has learned from the training data. However, as \citet{hutchinson2022underspecification} point out, ``Descriptions and depictions necessarily convey incomplete information about all but the most trivial scene.'' The model's approach to handling under-specification will therefore have varied and wide-ranging effects.

Thus, our research question is as follows: If we prompt the model to generate a person with particular social traits (as defined by the ABC Model), will the resulting images show the stereotypical demographic characteristics associated with those traits? We investigate this question using the 16 traits of the ABC Model and the demographic characteristics of skin colour, gender, and age, with three popular text-to-image models: DALL-E 2, Midjourney, and Stable Diffusion. We find that while not all traits generate stereotypical images, each model shows idiosyncratic biases along certain dimensions. We also observe intersectional biases, in particular a bias in all three systems associating the adjective of ``poor'' with darker-skinned males.

\section{Related Work}

In recent years, the machine learning research community has devoted significant effort to combating bias-related issues in computer vision applications. One line of work has analyzed the biases that stem from unbalanced class distributions in training datasets, resulting in systematic errors and poor performance on the minority classes. For example, as \citet{buolamwini2018gender} reported, the over-representation of light-skinned individuals in commonly-used facial recognition datasets leads to drastically larger error rates of commercial gender classification systems for darker-skinned females. Data augmentation methods can improve data balance, the efficacy of which is often measured by overall accuracy as well as more uniform performance across attributes such as race and gender \citep{deviyani2022assessing}. In a recent work, \citet{mitchell2020diversity} introduced quantitative metrics to directly measure the diversity and inclusion of a dataset, defining these concepts with respect to sociopolitical power differentials (gender, race, etc.) in management and organization sciences. Other works studied biases originated from annotations, such as linguistic biases, stereotypical descriptions, and unwarranted inferences about people's demographic traits in crowd-sourced annotations \citep{van2016stereotyping}, or reported bias when annotators make implicit decisions about what is worth mentioning in the annotations \citep{misra2016seeing}.


Besides data and annotation distributions, the choice of models can impact the fairness of trained algorithms. For example, computer vision models trained with zero-shot natural language supervision exhibit unexpected systematic errors associated with gender, race, and age traits, for specific design choices \citep{agarwal2021evaluating}. Also, image captioning models that learn to use contextual cues often exaggerate stereotypical cues present in the context to predict demographic traits \citep{hendricks2018women}. These observations call for task-specific and safety-focused evaluations to audit computer vision models for biased outcomes before deployment. \citet{raji2020saving} identified multiple ethical concerns in auditing commercial face recognition systems and recommended deliberate fairness evaluations as minimizing biases for some groups might cause unintended harms for others. 

Recently, work has begun that focuses on particular biases that have emerged in multi-modal language--vision machine learning systems. \citet{wolfe2022american} reported that racial biases about American identity, previously observed in social psychology, are learned by multi-modal embedding models and propagated to downstream tasks. Other works proposed evaluation frameworks to assess biases in text-to-image systems \citep{cho2022dall} or training mechanisms such as adversarial learning to reduce representation biases in language--vision models \citep{berg2022prompt}. Further, in a position paper, \citet{hutchinson2022underspecification} discussed social bias amplification, among other ethical concerns that arise from the use of text-to-image systems. They identified ambiguity and under-specification as the root causes of these risks and proposed conceptual frameworks to manage them. Specifically, they introduced two approaches to deal with under-specification: \textit{Ambiguity In, Ambiguity Out}
(AIAO) and \textit{Ambiguity In, Diversity Out} (AIDO). In the AIAO approach, the model is encouraged to generate ambiguous images when the concepts in input text are under-specified. In the alternative approach, AIDO, the preferable behaviour is generating a set of maximally diverse images to cover the space of possibilities for the under-specified concept. We consider both of these approaches in our current analyses.

\begin{table*}[]
    \centering
    \begin{tabular}{m{1em}c|m{1em}c|m{1em}c}
    \hline
    \multicolumn{1}{c}{}&\multicolumn{1}{c}{}&\multicolumn{1}{c}{}&\multicolumn{1}{c}{}&\multicolumn{1}{c}{}&\\[-0.75em]
\rotatebox{90}{\textbf{Agency} }&
\makecell{powerless $\leftrightarrow$ powerful \\ low-status $\leftrightarrow$ high-status\\ dominated $\leftrightarrow$ dominating \\ poor $\leftrightarrow$ wealthy \\ meek $\leftrightarrow$ confident \\ passive $\leftrightarrow$ competitive}&
\rotatebox{90}{\textbf{Belief} }&
\makecell{religious $\leftrightarrow$ science-oriented \\ conventional $\leftrightarrow$ alternative\\  conservative $\leftrightarrow$ liberal\\traditional $\leftrightarrow$ modern  }&
\rotatebox{90}{\textbf{Communion} }&
\makecell{untrustworthy $\leftrightarrow$ trustworthy \\ dishonest $\leftrightarrow$ sincere \\unfriendly $\leftrightarrow$ friendly \\threatening $\leftrightarrow$ benevolent \\ unpleasant $\leftrightarrow$ likable\\egoistic $\leftrightarrow$ altruistic }\\
\multicolumn{1}{c}{}&\multicolumn{1}{c}{}&\multicolumn{1}{c}{}&\multicolumn{1}{c}{}&\multicolumn{1}{c}{}&\\[-0.75em]
    \hline
    \end{tabular}
    \caption{List of stereotype dimensions and corresponding traits in the ABC model, adapted from  \citet{cao2022theory}.}
    \label{tab:adjectives}
\end{table*}

\subsection{Social Stereotypes}

We draw our hypotheses from the existing survey-based literature on prevalent stereotypes in North American society. In the paper introducing the ABC model, \citet{koch2016abc} present results that place various social groups in the three-dimensional Agency-Beliefs-Communion space.
Most relevant to our work here are social groups defined by gender, age, or skin colour. While the specific results vary somewhat across their sub-studies, some consistent patterns are seen: in terms of Agency, white people are rated higher than people of colour, old or elderly people are rated higher than young people, and men are rated higher than women. In terms of Beliefs, young people are rated as more progressive than older people, and there is no obvious distinction based on gender or skin colour. In terms of Communion, white people are rated higher than Black people, although Latinos, Hispanics, Asians, and Indians are also rated as positive-communion. Older people are also seen as higher on this dimension than younger people. The experimental design does not directly compare communion values for men and women. However, other related literature confirms many of these predictions and also reports that women are seen as more Communal (warm, friendly) than men \citep{fiske2018model, nicolas2022spontaneous}, and white people are seen as more modern and science-oriented (high-Beliefs) than Black people \citep{cao2022theory}.

As a result, our hypotheses are as follows. Given the under-specification of our prompts with respect to the demographic characteristics of the generated image subject, the text-to-image models will default to social stereotypes learned from the data, namely: 

\begin{itemize}
    \item \textbf{High-agency} words will tend to generate images of people with  lighter skin, older age, and  male gender, while \textbf{low-agency} words will tend to generate images of people with darker skin, younger age, and female gender.
    \item \textbf{High-belief} (progressive) words will tend to generate images of younger and lighter-skinned people, while \textbf{low-belief} (conservative) words will tend to generate images of older and darker-skinned people.
    \item \textbf{High-communion} words will tend to generate images of people with lighter skin, older age, and female gender, while \textbf{low-communion} words will tend to generate images of people with darker skin, younger age, and male gender.
\end{itemize}

\section{Methodology}

We first describe the three contemporary text-to-image systems evaluated in the current study, and then provide details on image generation and annotation processes.

\subsection{Text-to-Image Systems}

All three systems evaluated in this study, DALL-E 2,\footnote{\url{https://openai.com/dall-e-2/}} Midjourney,\footnote{\url{https://www.midjourney.com}} and Stable Duffusion,\footnote{\url{https://huggingface.co/spaces/stabilityai/stable-diffusion}} generate original images from textual prompts and/or uploaded images. They are based on state-of-the-art image generation technology, like diffusion models \cite{sohl2015deep,nichol2021glide} and CLIP image embeddings \cite{radford2021learning}, and are trained on millions and billions of  text--image examples scraped from the web. We briefly discuss each system and provide more details in the Appendix. All images used in the study were generated in October 2022. 

\noindent{\textbf{DALL-E 2:}} This is a research and production system released as beta version by OpenAI in July 2022. 
DALL-E 2 (hereafter, simply `DALL-E') aims to create photorealistic, diverse images, that closely represent the textual prompts \cite{ramesh2022hierarchical}.   
To more accurately reflect the diversity of the world’s population and to prevent the dissemination of harmful stereotypes, the system has been extended to 
further diversify its output for under-specified prompts of portraying a person (e.g., `a portrait of a teacher').\footnote{\url{https://openai.com/blog/reducing-bias-and-improving-safety-in-dall-e-2/}}

\noindent{\textbf{Midjourney:}} This system was created by an independent research lab Midjourney and released as beta version in July 2022; we used the most recent version, v3.    
It has been designed as a social app where users generate images along side other users in public community channels through the chat service Discord. 
The system details have not been publicly released. 

\noindent{\textbf{Stable Diffusion:}} This system was publicly released by Stability AI under a Creative ML OpenRAIL-M license 
  in August 2022. 
It is based on latent diffusion model by
\citet{Rombach_2022_CVPR}. 
The system was trained to produce aesthetically pleasing images using LAION-Aesthetics dataset. 
We accessed Stable Diffusion v1.5 through the DreamStudio API with default settings.

\subsection{Image Generation}

For each of the three systems, we used templates to produce prompts containing each of the adjectives in Table~\ref{tab:adjectives}. These adjectives were taken from \citet{koch2016abc} with a few minor variations: the original ABC Model uses the adjectives \textit{warm}, \textit{cold}, and \textit{repellent}, which we found to be too semantically ambiguous to produce reliable results (e.g., would generate a person who was physically freezing cold). These words were replaced with \textit{friendly}, \textit{unfriendly}, and \textit{unpleasant}, respectively. \citet{koch2016abc} also use two words which have extremely low frequency (less than one instance per million in the SUBTLEX-US corpus), namely \textit{unconfident} and \textit{unassertive}; these were replaced with \textit{meek} and \textit{passive}. 

Our basic prompt took the form of: \texttt{portrait of a $\langle$adjective$\rangle$ person}. The word \texttt{portrait} cues the system to show the face of the subject, which contains most of the visual information needed to make the demographic  annotations. For each model, we found that we needed to adapt the prompt slightly to achieve acceptable results. 
For DALL-E, the prompt as written was very effective in generating colour, photo-realistic results; in the few cases that a drawing or black-and-white image were generated, we re-ran the generation until a colour, photo-realistic result was achieved. 
For Midjourney, the basic prompt had a tendency to generate more artistic interpretations. Adding the flag \texttt{--testp} helped generate photo-realistic results, but they were mostly black-and-white. Adding the keywords \texttt{color photograph} was not effective, as it generated highly stylized colours that obscured the skin colour of the subjects. Instead, we found the keywords \texttt{Kodak Portra 400} (a popular colour film stock) to be highly effective at producing colour, photorealistic results. Similarly for Stable Diffusion, adding the \texttt{Kodak Portra 400} keywords to the end of the prompt led to the generation of interpretable results as opposed to painterly, abstract images.

Since DALL-E outputs images in batches of 4, we decided to generate 24 images per trait (12 per pole of each trait). We additionally generated 24 baseline images for each model, using the basic prompt of \texttt{portrait of a person}, with no trait adjective. Thus for each of the three models, we generated a total of 408 images. 

\subsection{Image Annotation}

Each image was annotated by three annotators (the authors of the paper). Our demographic characteristics of interest were gender, skin colour, and age. The process of inferring demographic characteristics from images has numerous ethical challenges. We outline our processes and assumptions here, with a more detailed discussion in the Appendix.

First, we emphasize that we are annotating perceived demographic characteristics of \textit{AI-generated images}, not real people. We are doing this with the goal of assessing the diversity of the outputs, not of categorizing real individuals. To that end, we have also decided not to make our annotations publicly available, so as not to facilitate such a use case.

Second, following best practices from the literature \cite{buolamwini2018gender}, we do not attempt to categorize particular races/ethnicities, but rather focus on the more objective measure of skin colour (from ``lighter'' to ``darker''). We also recognize gender as being a non-binary variable and allow for a gender-neutral annotation. 

Finally, we combine the annotations using an averaging technique such that each annotator's judgement is equally weighted, rather than using a majority-voting scheme. 
The full annotation instructions are available in the Appendix. Briefly, each demographic variable can receive one of four possible annotations (gender: male, female, gender neutral, or no gender information available; skin colour: darker, lighter, in-between, or no skin colour information available; age: older, younger, in-between, or no age information available). These categorical annotations are converted to numerical values and averaged over the three annotators. As a concrete example, if two annotators marked an image subject as having darker skin (+1) and one annotated it as in-between (0), then the image would be assigned a summary skin colour value of 0.67.

\section{Results}

\subsection{Annotation Reliability}

The Cohen's Kappa values for inter-annotator agreement are given in Table~\ref{tab:kappa}. In general, the gender annotation had highest agreement, and the skin colour annotation had the lowest. This can be partly attributed to the fact that perceived gender typically fell into either `male' or `female' categories, with only a few gender-ambiguous images in between, while skin colour ranged across a full spectrum, creating more annotation uncertainty between categories. (Although, note that this issue is partially alleviated by our annotation averaging technique.) Furthermore, skin colour estimation was confounded by varying lighting conditions in the images. The agreement values are overall lower for the Stable Diffusion dataset, reflecting a qualitatively lower degree of photorealism in the generated images. 

\begin{table}[]
    \centering
    \begin{tabular}{rr c c c}
    \hline
    Dataset & Gender & Skin Colour & Age \\
    \hline
    Midjourney & 0.84 & 0.57 & 0.76 \\
    DALL-E     & 0.92 & 0.64 & 0.68 \\
    Stable Diffusion & 0.75  & 0.54 & 0.54 \\
    \hline 
    \end{tabular}
    \caption{Cohen's Kappa ($\kappa$) metric of inter-annotator agreement for each demographic variable and each dataset.}
    \label{tab:kappa}
\end{table}

\subsection{Ambiguity In, Ambiguity Out}

As mentioned above, one viable strategy for dealing with ambiguous or under-specified inputs is to in turn produce ambiguous or under-specified outputs (AIAO). Our experimental paradigm constrained the systems' ability to deploy this strategy, by first prompting for a portrait (implying that the face should be visible), by constraining the analysis to colour images, and by prompting for photographic-style results. Loosening these constraints would no doubt lead to more creative and interpretative outputs by the models. Nonetheless, in some cases the generated images were ambiguous with respect to one or more of the demographic variables. Common methods of creating ambiguity included: positioning the subject facing away from the camera, obscuring the subject's face with an object, generating non-realistic skin colours (e.g., purple), or blurring the subject's features. 
The rate at which each model produced such AIAO images is given in Table~\ref{tab:AIAO}. In the following sections, we remove these images from the analysis and focus on the alternate strategy of Ambiguity In, Diversity Out (AIDO).

\begin{table}[]
    \centering
    \begin{tabular}{r c c c }
    \hline
    Dataset & Gender & Skin Colour & Age  \\
    \hline 
 Midjourney & 0.06 & 0.06 & 0.08 \\
  DALL-E  & 0.00    & 0.03 & 0.04 \\ 
  Stable Diffusion & 0.06 & 0.01 & 0.11 \\
  \hline 
    \end{tabular}
    \caption{Proportion of images for which at least one annotator indicated that no visual cues to the given demographic variable were present in the image, reflecting an `Ambiguity In, Ambiguity Out' (AIAO) strategy. }
    \label{tab:AIAO}
\end{table}

\subsection{Baseline Results}

\begin{figure}
    \centering
    \includegraphics[width=\columnwidth]{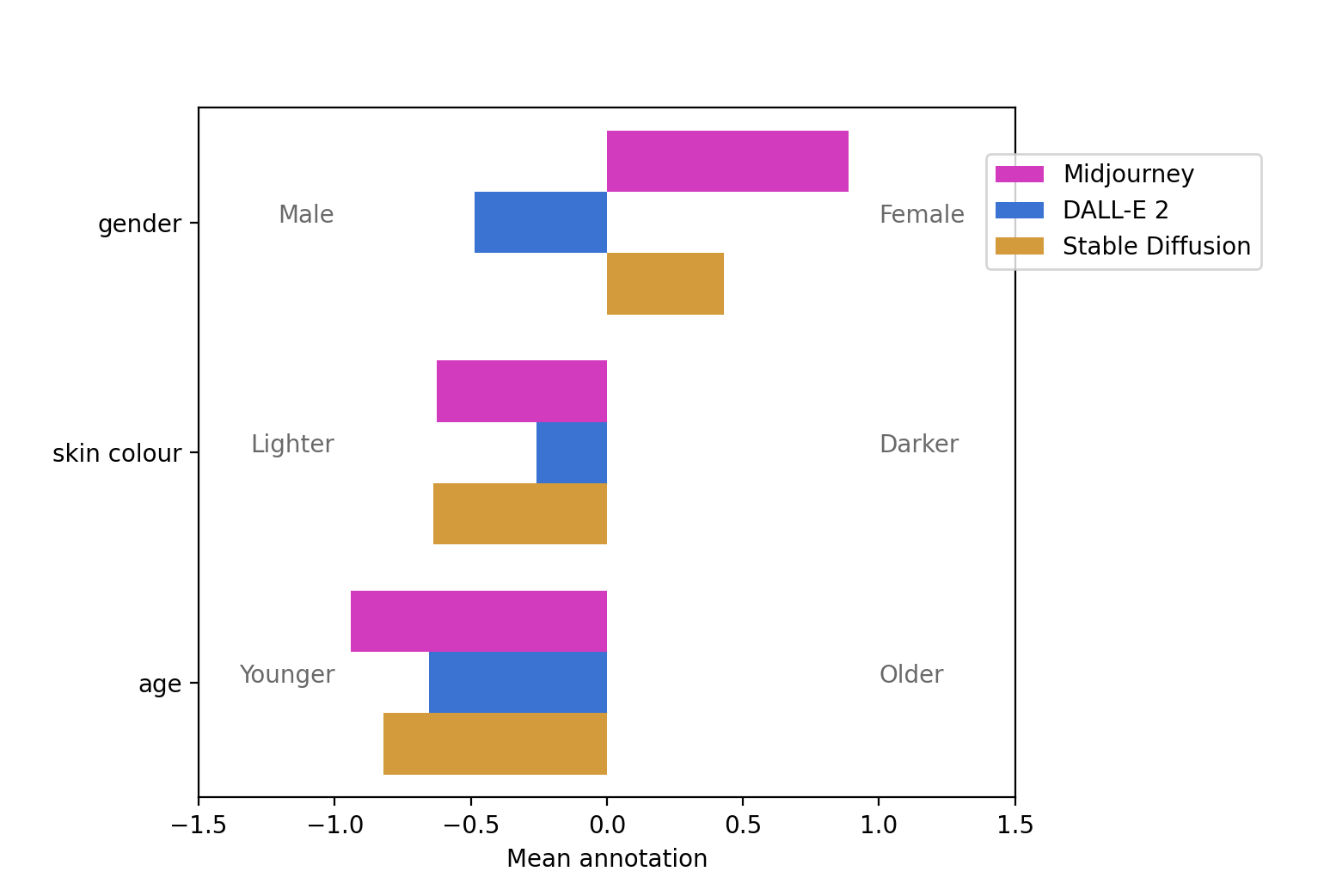}
    \caption{Baseline values for gender, skin colour, and age for the three models. 
    }
    \label{fig:baseline}
\end{figure}

Figure~\ref{fig:baseline} shows the baseline results of prompting each model for \texttt{portrait of a person}. Midjourney and Stable Diffusion have a tendency to produce female subjects, while DALL-E tends to produce males. All three systems have a tendency to generate subjects with lighter skin tones, although DALL-E less so than the other two. All three systems also have a strong tendency to produce younger-looking subjects. 

\subsection{ABC Model Results}

\begin{figure*}[t]
     \centering
     \begin{subfigure}[b]{0.3\textwidth}
         \centering
         \includegraphics[width=\textwidth]{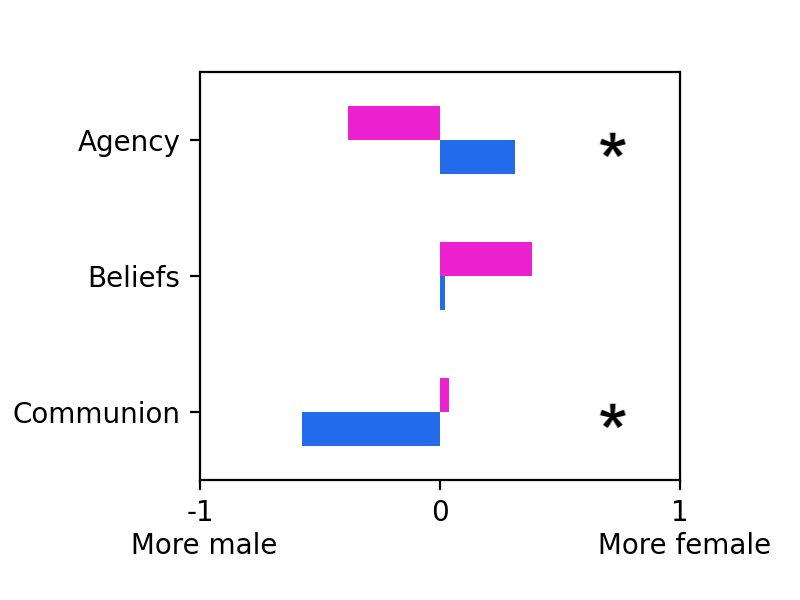}
         \caption{Midjourney -- Gender}
         \label{fig:gender_MJ}
     \end{subfigure}
     \hfill
     \begin{subfigure}[b]{0.3\textwidth}
         \centering
         \includegraphics[width=\textwidth]{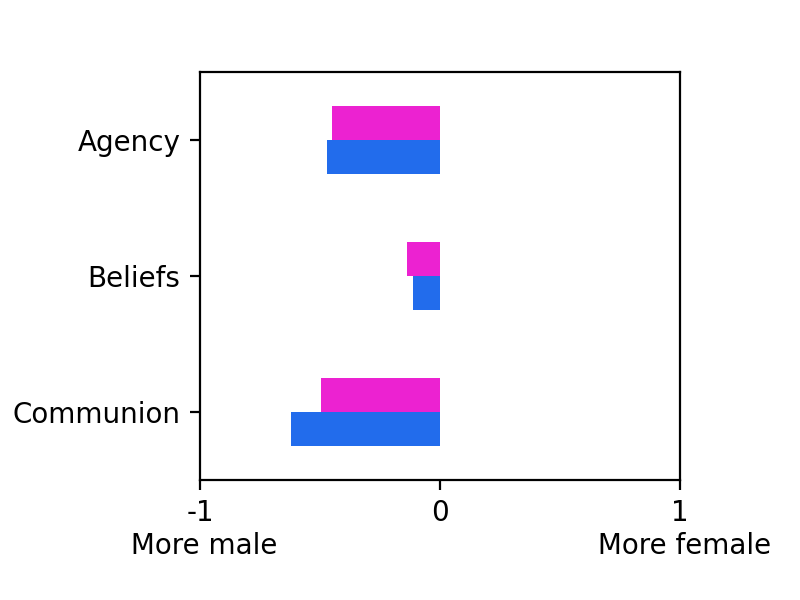}
         \caption{DALL-E -- Gender}
         \label{fig:gender_DALLE}
     \end{subfigure}
     \hfill
     \begin{subfigure}[b]{0.3\textwidth}
         \centering
         \includegraphics[width=\textwidth]{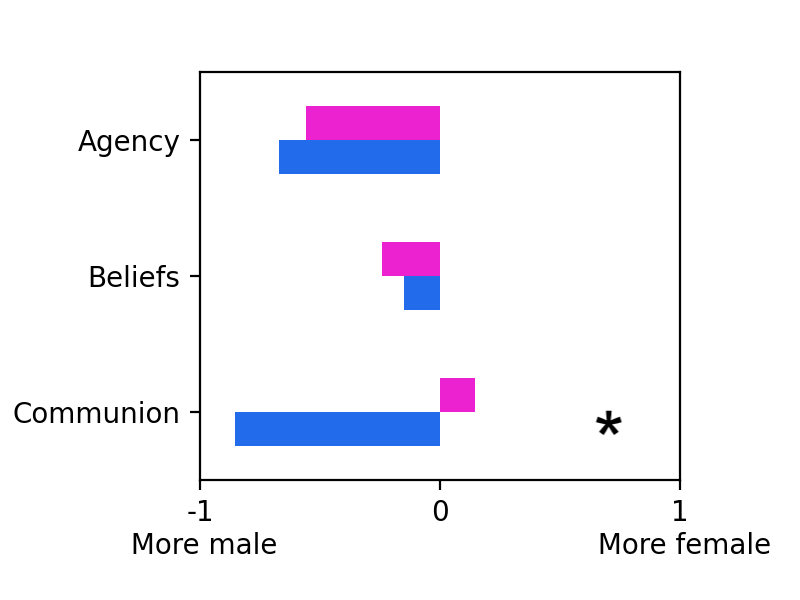}
         \caption{Stable Diffusion -- Gender}
         \label{fig:gender_SD}
     \end{subfigure}
 
      \begin{subfigure}[b]{0.3\textwidth}
         \centering
         \includegraphics[width=\textwidth]{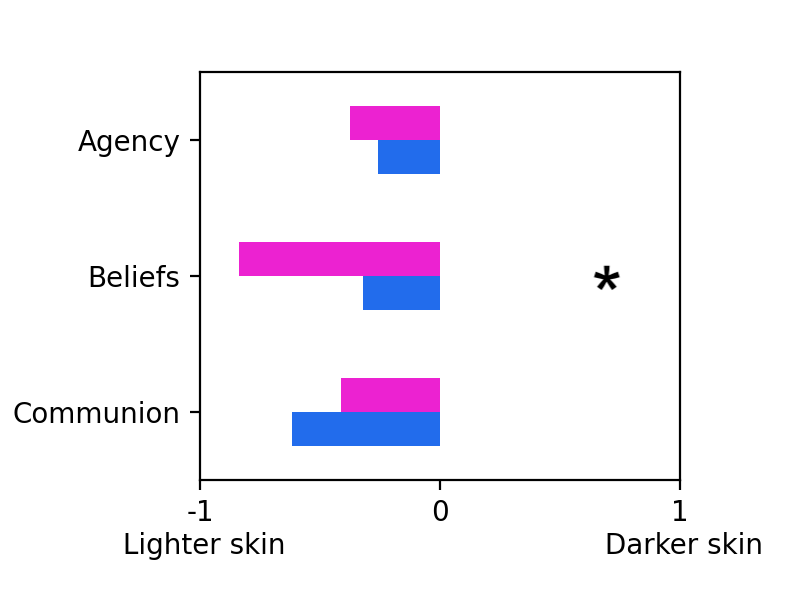}
         \caption{Midjourney -- Skin Colour}
         \label{fig:race_MJ}
     \end{subfigure}
     \hfill
     \begin{subfigure}[b]{0.3\textwidth}
         \centering
         \includegraphics[width=\textwidth]{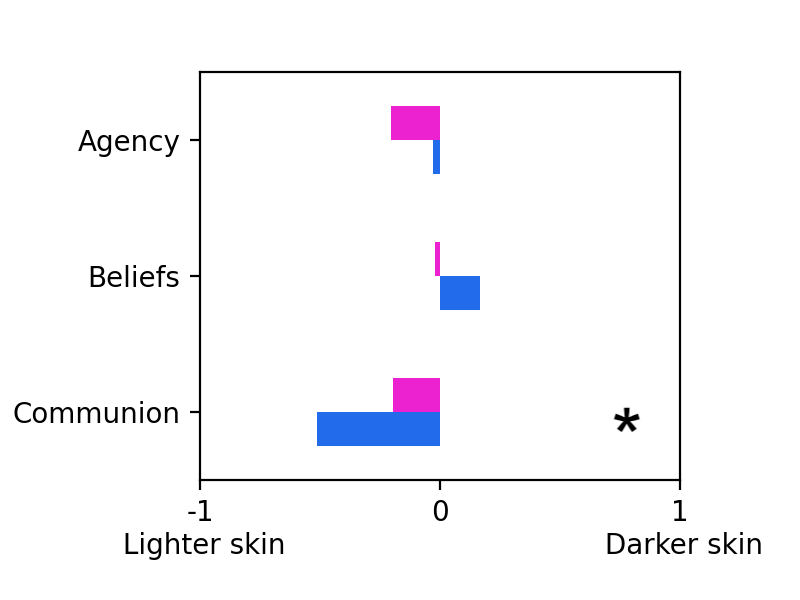}
         \caption{DALL-E -- Skin Colour}
         \label{fig:race_DALLE}
     \end{subfigure}
     \hfill
     \begin{subfigure}[b]{0.3\textwidth}
         \centering
         \includegraphics[width=\textwidth]{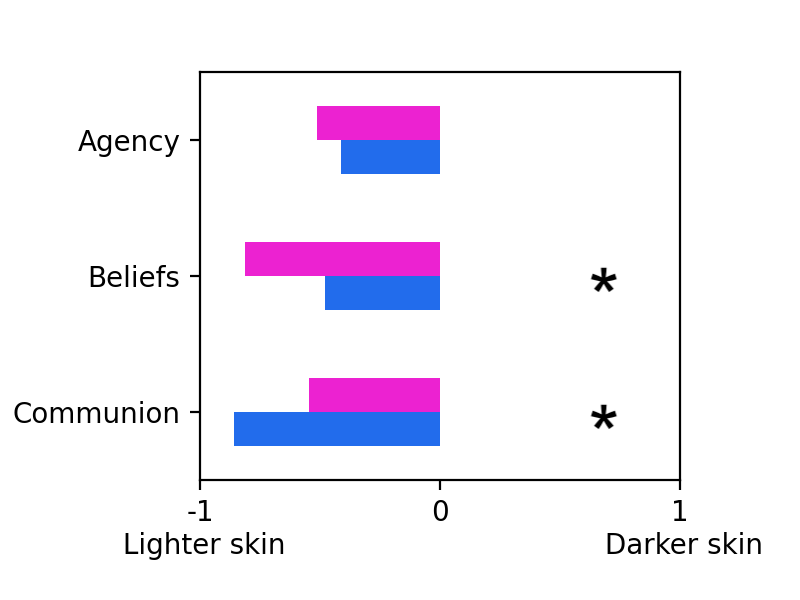}
         \caption{Stable Diffusion -- Skin Colour}
         \label{fig:race_SD}
     \end{subfigure}
 
       \begin{subfigure}[b]{0.3\textwidth}
         \centering
         \includegraphics[width=\textwidth]{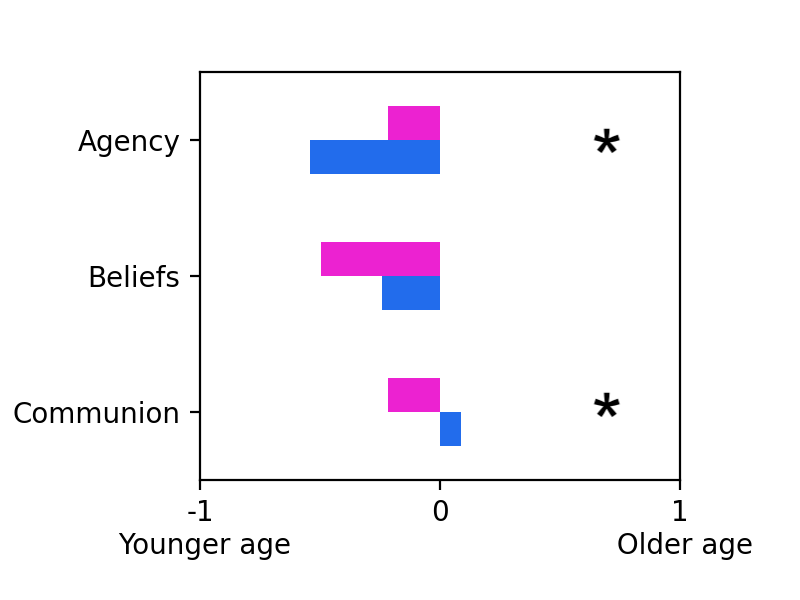}
         \caption{Midjourney -- Age}
         \label{fig:age_MJ}
     \end{subfigure}
     \hfill
     \begin{subfigure}[b]{0.3\textwidth}
         \centering
         \includegraphics[width=\textwidth]{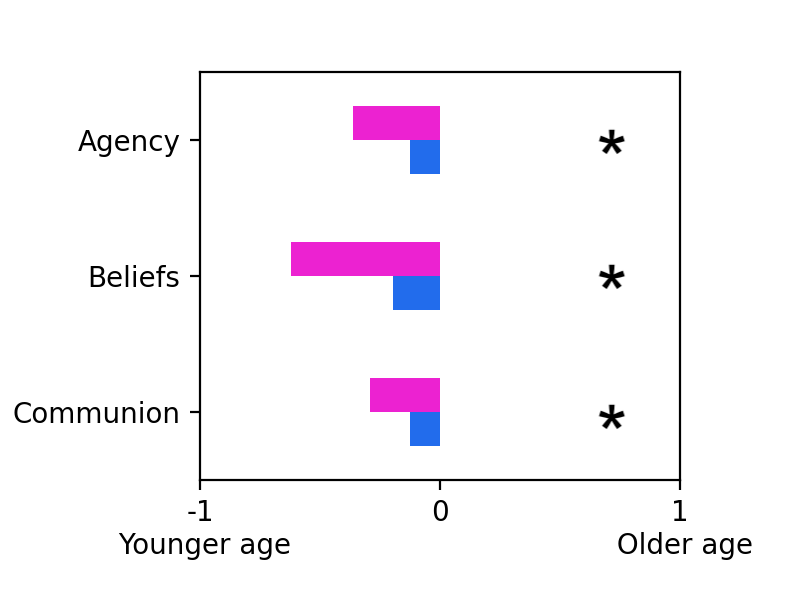}
         \caption{DALL-E -- Age}
         \label{fig:age_DALLE}
     \end{subfigure}
     \hfill
     \begin{subfigure}[b]{0.3\textwidth}
         \centering
         \includegraphics[width=\textwidth]{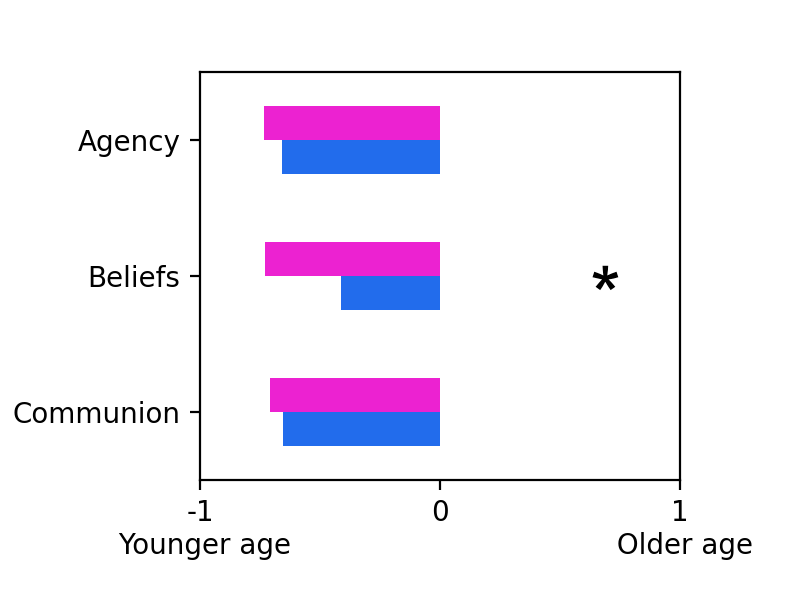}
         \caption{Stable Diffusion -- Age}
         \label{fig:age_SD}
     \end{subfigure}

        \caption{Average annotation values for gender, skin colour, and age, for each of the ABC dimensions and for each model. The positive pole of each trait is shown in pink; the negative pole is shown in blue. Significant differences between positive and negative traits are indicated with an asterisk.}
        \label{fig:ABC}
\end{figure*}

Figure~\ref{fig:ABC} shows, for each demographic variable, the average annotation across all images associated with the positive and negative poles in each ABC dimension. That is, for Agency, the positive pole includes all images generated from \textit{powerful}, \textit{dominating}, \textit{high-status}, etc., and the negative pole includes all images generated from \textit{powerless}, \textit{dominated}, \textit{low-status}, etc., and similarly for Beliefs and Communion. Thus for Agency, each of the positive and negative values are averaged over 3 annotators per image $\times$ 12 images per trait $\times$ 6 traits, or 216 annotations. When there is a significant difference between the positive and negative poles, this is indicated with an asterisk in the figure ($p < 0.05$, Mann-Whitney U-Test). This  comparison corresponds to the concept of stereotype \textbf{\textit{direction}} as defined by \citet{nicolas2022spontaneous}: ``Group X is associated with positive or negative direction on a given trait.'' However, that paper also introduces the complementary notion of \textbf{\textit{representativeness}}, i.e., Group X being highly associated with a given trait, regardless of polarity. We discuss the results with reference to both of these concepts in the following.

Beginning with the gender variable, in Fig.~\ref{fig:gender_MJ} we observe a significant difference in the Midjourney results for Agency, with high-agency words more likely to generate images representing male gender, and low-agency words more likely to generate images representing female gender, as hypothesized. There is also a significant difference in Communion, with low-communion words more associated with male gender as expected. When we break these dimensions down by trait (shown in the Appendix), this corresponds to more images of men generated for words like \textit{high-status} and \textit{dominating} (high-Agency) and \textit{dishonest} and \textit{unpleasant} (low-Communion), while more images of women were generated for adjectives like \textit{powerless}, \textit{friendly}, and \textit{likable}.
In the case of DALL-E (Fig.~\ref{fig:gender_DALLE}), there is no significant difference on gender along any of the three dimensions. However, we do observe a  difference in \textit{representativeness}: specifically, that DALL-E has a tendency to produce more males than females for all ABC dimensions, regardless of polarity. 
For Stable Diffusion (Fig.~\ref{fig:gender_SD}), similar to Midjourney we observe that low-communion words are significantly associated with male gender, and similar to DALL-E that the male gender is over-represented in general. 

Turning now to the variable of skin colour, for Midjourney (Fig.~\ref{fig:race_MJ}) we observe a significant difference along the Beliefs dimension, with progressive beliefs more highly-associated with lighter skin colours, as expected. 
This trend is driven by a high proportion of lighter-skinned subjects generated for the prompts \textit{science-oriented} (versus \textit{religious}) and \textit{modern} (versus \textit{traditional}). In terms of representativeness, all dimensions have a tendency towards lighter skin. 
For DALL-E (Fig.~\ref{fig:race_DALLE}), for Agency and Beliefs we see no significant difference in direction with respect to skin colour, and more equal representation. However, there is a significant difference in direction for Communion, with low-communion words more associated with lighter skin, in contradiction to our hypothesis. This trend is driven by adjectives \textit{untrustworthy}, \textit{threatening}, and \textit{unpleasant}.
In the case of Stable Diffusion (Fig.~\ref{fig:race_SD}), we again observe an over-representation of images depicting lighter-skinned subjects in all dimensions. Additionally, the system shows a significant difference in the dimensions of Beliefs (progressive beliefs more associated with lighter skin) and Communion (low communion more associated with lighter skin, specifically for the words  \textit{dishonest} and \textit{threatening}).

Finally, considering age: Midjourney (Fig.~\ref{fig:age_MJ}) shows a significant difference in Agency and Communion, with low-agency words  and high-communion words more associated with images of younger people (recall, the first trend is in keeping with our hypotheses but the second is not).
For DALL-E (Fig.~\ref{fig:age_DALLE}), the positive poles of all three ABC dimensions are associated with younger age, and this difference is significant in all cases. For Stable Diffusion (Fig~\ref{fig:age_SD}), the same trend occurs, although it is only significant in the Beliefs dimension. However, for Stable Diffusion we also note a highly-skewed representativeness towards younger age in all dimensions. In particular, for all three systems, the adjective \textit{likable} was highly associated with younger age, with its contrast adjective \textit{unpleasant} ranging from moderately to highly associated with older age.

\subsection{Intersectional Results}

\begin{figure}
     \centering
     \begin{subfigure}[b]{0.45\textwidth}
         \centering
         \includegraphics[width=\textwidth]{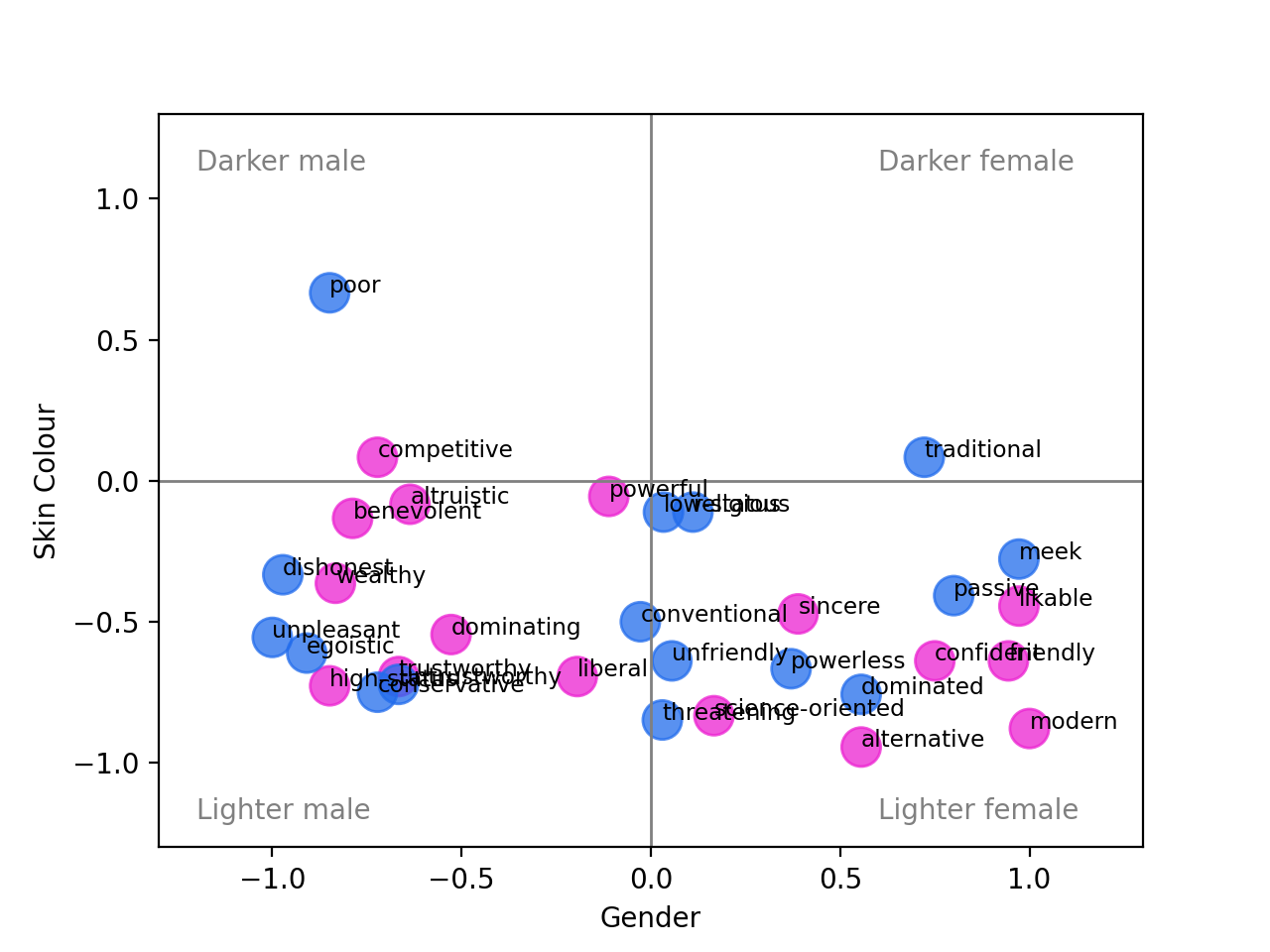}
         \caption{Midjourney}
         \label{fig:MJ_race_gender}
     \end{subfigure}
     \hfill
     \begin{subfigure}[b]{0.45\textwidth}
         \centering
         \includegraphics[width=\textwidth]{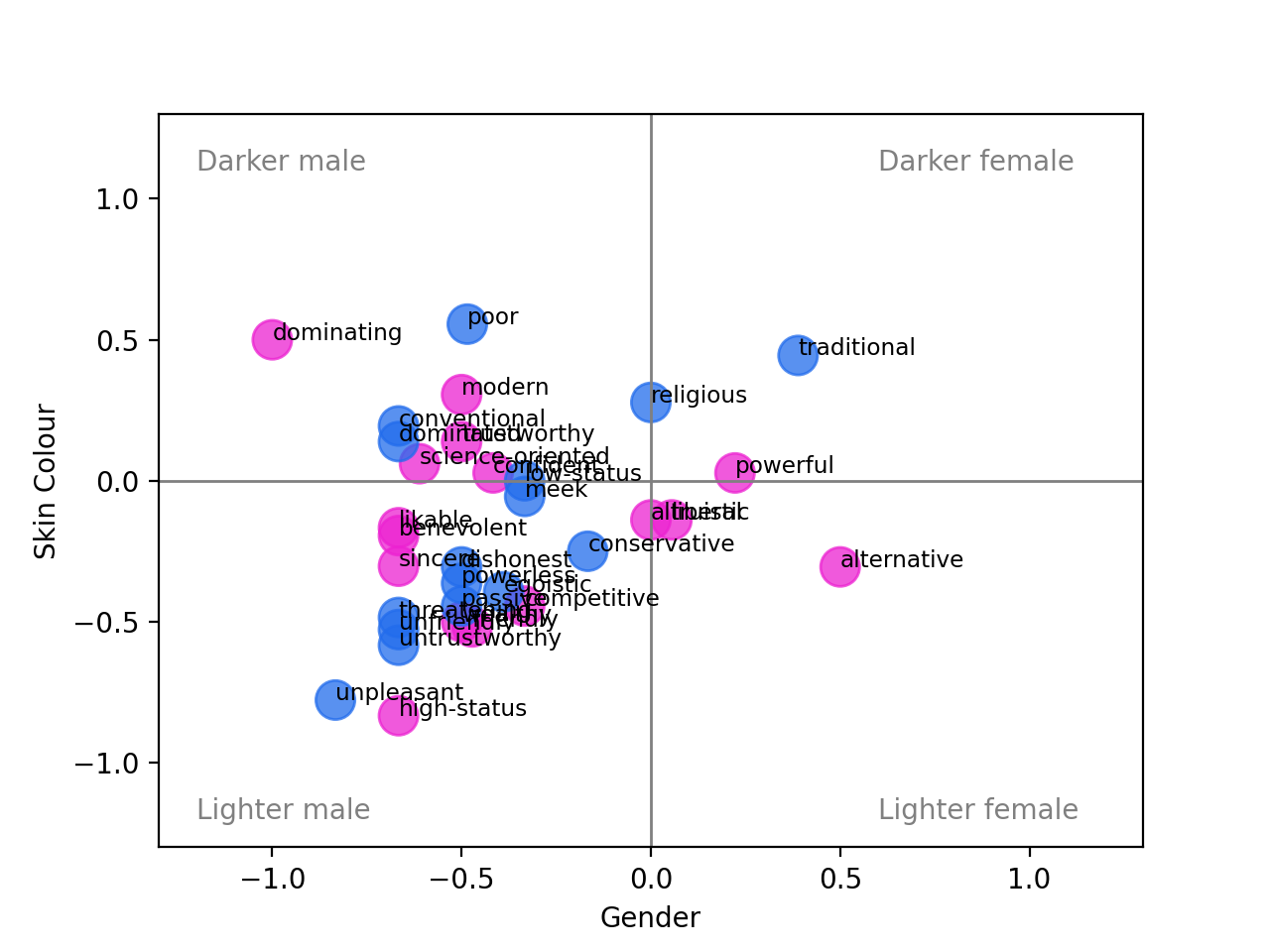}
         \caption{DALL-E}
         \label{fig:DALLE_race_gender}
     \end{subfigure}
     \hfill
     \begin{subfigure}[b]{0.45\textwidth}
         \centering
         \includegraphics[width=\textwidth]{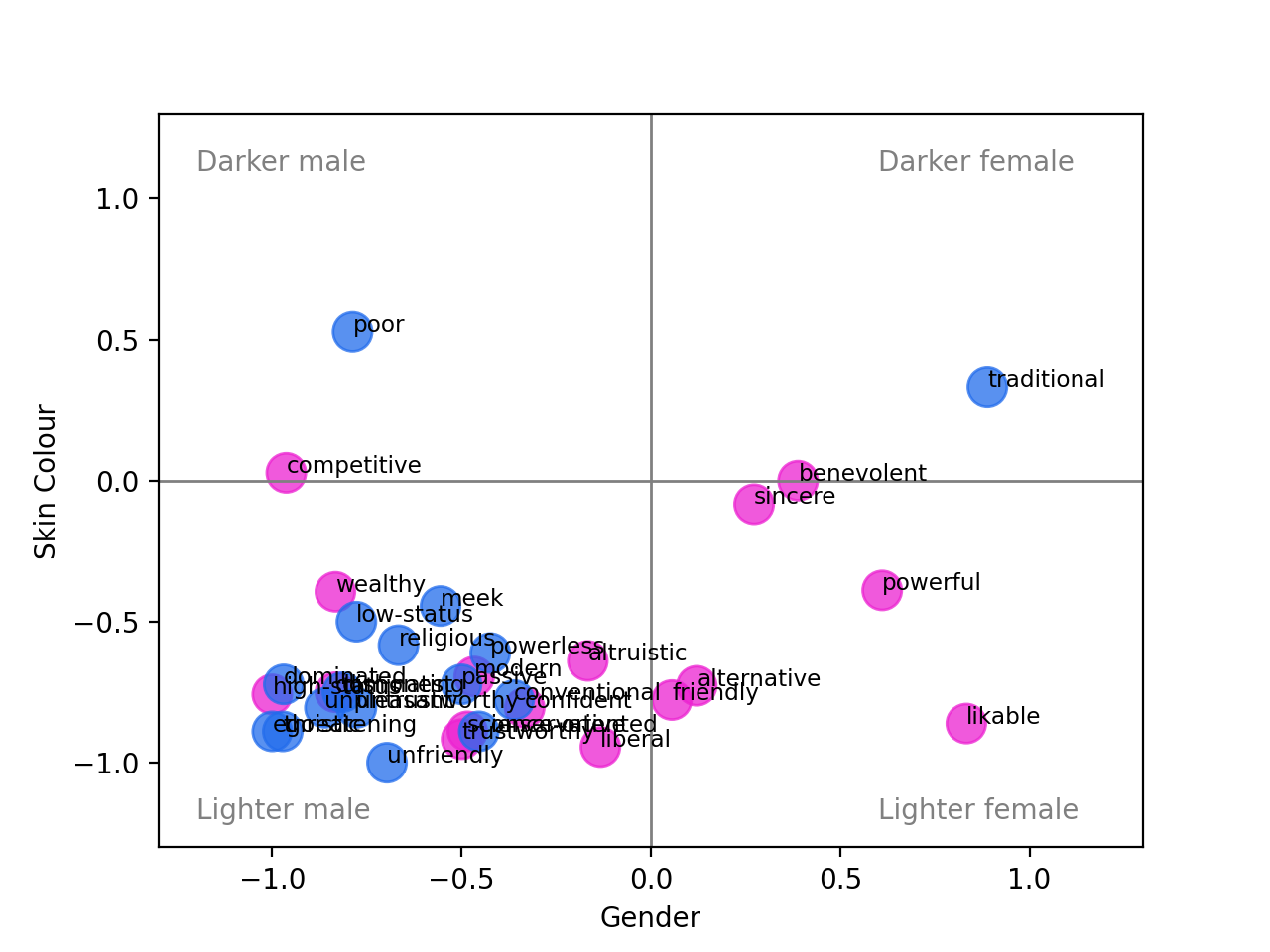}
         \caption{Stable Diffusion}
         \label{fig:SD_race_gender}
     \end{subfigure}

    \caption{Intersectional view of each trait (skin colour x gender). The positive poles of each trait are shown in pink; the negative poles are shown in blue.}
    \label{fig:race_gender}
\end{figure}

While looking at each dimension individually is informative, additional insight can be obtained by considering the results intersectionally. Figure~\ref{fig:race_gender} shows the average skin colour and gender annotation for each pole of each trait.  
In Figure~\ref{fig:MJ_race_gender} it is obvious that while Midjourney generates images across the range of genders, the only case where the model had a strong tendency to generate males with darker skin was for the adjective \textit{poor}. It had a slight tendency to generate darker-skinned males for \textit{competitive} (with images typically showing athletes), and a slight tendency to generate darker-skinned women for \textit{traditional} (with images showing women in ``traditional'' dress from various cultures). 

For DALL-E (Fig.~\ref{fig:DALLE_race_gender}), we observe a different distribution. DALL-E is more likely to generate both darker- and lighter-skinned people, but those people tend to be male. Images of lighter-skinned women were generated for the adjective \textit{alternative}, and images of darker-skinned women were generated for \textit{traditional}, as above. While the distribution across skin colour is more equitable in general, we do again note that the adjective \textit{poor} tends to generate darker-skinned males. 

Turning to the Stable Diffusion results in Figure~\ref{fig:SD_race_gender}, we see some similarities and some differences in comparison with the other models. Once again, darker skinned males are generated mostly by the adjective \textit{poor}, with a slight trend for \textit{competitive}, and again darker-skinned females are associated with \textit{traditional}. Unlike DALL-E, there is a higher density of points in the lighter-skinned female quadrant, and unlike Midjourney the points all tend to be associated with `positive' adjectives: \textit{benevolent} and \textit{sincere} occur right along the 0 axis for skin colour, with \textit{powerful} and \textit{likable} associated with lighter-skinned women.

For lack of space, the corresponding figures plotting age versus skin colour and age versus gender are given in the Appendix, but we briefly summarize the findings here. Figure~\ref{fig:race_age} shows that the adjective \textit{poor} is also anomalous when we consider age versus skin colour: in the case of both DALL-E and Stable Diffusion, it is the point closest to the upper right (oldest and darkest-skinned). Overall, Midjourney outputs a range of ages, but primarily lighter skin colours (points concentrated on the left-most quadrants), DALL-E produces a range of skin colours but mostly younger faces (points concentrated in the bottom two quadrants), and Stable Diffusion produces mostly lighter, younger faces (points mostly in the bottom, left quadrant).

When we consider age versus gender (Fig.~\ref{fig:gender_age}), Midjourney shows a surprisingly linear negative trend, with some adjectives associated with older males, and others associated with younger females, but no traits associated primarily with older females, and only one trait (\textit{competitive}) associated primarily with younger males. DALL-E and Stable Diffusion both exhibit a trend of generating younger males.

\section{Summary and Discussion}

Many of the significant differences 
in Figure~\ref{fig:ABC} did confirm our hypotheses: high-Agency words generated  more men than women, as did low-Communion words, and in the case of Midjourney, low-Agency words were associated with younger age. As well, both Midjourney and Stable Diffusion showed a significant tendency to associate progressive Beliefs with lighter skin, primarily driven by the traits \textit{modern--traditional} and \textit{science-oriented--religious}, as also reported by \citep{cao2022theory}. 
In contrast to our hypotheses, lighter skin was associated with low-communion adjectives for both DALL-E and Stable Diffusion.  
We also found an unexpected trend of high-communion associated with younger age. 
Combined with the fact that all three models showed a preference for generating more images of younger people, it appears that age-based bias may need to be addressed by the developers of such systems. 

Considering the intersectional results, the lack of representation of darker females is particularly striking, and unfortunately not surprising, given previous scholarship on intersectional bias in AI systems \citep{d2020data}. That all three systems also associated poverty with dark-skinned males warrants further investigation. Examples of the images produced for prompts \textit{poor} and \textit{wealthy} are given in the Appendix, Figure~\ref{fig:ex_race_gender}.

These results must be considered preliminary, due to the small sample size and limited variety in the prompt structure. However, we offer a brief discussion on some of the factors that may contribute to the differences seen across the three models, as well as differences with the ABC Model.

First, we must consider the effect of how the training data was collected: largely by scraping the web for image-caption pairs. \citet{hutchinson2022underspecification} describe the many different relationships between captions and images that can exist, ranging from simple descriptions of the visual content of the image, to providing interpretations, explanations, or additional information. \citet{misra2016seeing} discuss the human reporting bias that exists in image datasets, as annotators choose what is ``worth mentioning'' in an image. In particular, traits which are seen as stereotype-consistent or ``default'' are often not mentioned; for example, \citet{van2016stereotyping} reports that the race of babies in the Flickr30k image dataset is not typically mentioned, unless the baby is Black or Asian. White babies are assumed to be the default. Hence, some of the stereotypical associations that we do \textit{not} see may potentially be because internet-users do not explicitly mention default traits in image captions. 

Another factor is the intended use case for these text-to-image systems. Midjourney and Stable Diffusion in particular have been marketed as methods of generating AI artwork. Therefore, the systems are optimized to produce ``aesthetically-pleasing'' results, with users providing feedback on what exactly that means to them. Historically, Western art has been biased towards images of the female form \citep{nead2002female}, as well as aesthetic preferences for youth and beauty. Therefore it is also possible that some of this aesthetic bias for what makes ``beautiful'' art is affecting the output distribution of these systems. 

Finally, it should be acknowledged that the human creators of such systems make normative design choices. OpenAI has published blog posts describing their efforts to reduce bias and improve safety in DALL-E 2, and in Figure~\ref{fig:ABC} we do see fewer extreme disparities in representation, particularly with respect to skin colour. On the other hand, expressing a philosophy of user freedom, Stable Diffusion has rejected the approach of filtering input and/or output content, and puts the responsibility on the user to use the system appropriately. The tension between freedom of expression and harm reduction has been seen in many areas of artificial intelligence, and continues to be an open -- and potentially unsolvable -- question. 

\section{Conclusion and Future Work}

Although not all systems showed the same types of stereotypical biases, each one demonstrated some room for improvement. In particular, we believe that analyzing age-related bias will be one fruitful area of research. This work also points to the need for further investigation of the relationships between race, gender, and economic status that have been encoded in such systems. Future work should involve confirming (or dis-confirming) the presence of such biases with bigger sample sizes and more varied prompt structure and content. 

One key open question is to what extent the bias originates from the training data, the model architecture, or the model parameters. The answer to that question will help inform appropriate de-biasing methods at training time. Another promising avenue of research involves mitigating bias at inference time through careful prompt engineering. For example, \citet{bansal2022well} report that modifying prompts with phrases such as `irrespective of gender' can encourage text-to-image models to generate outputs of various genders.  The further development of such intervention strategies will also help improve the fairness and diversity of model outputs.

\bibliography{aaai23.bib}  

\clearpage 
\newpage 

\appendix 


\section{APPENDIX}

\section{Ethical Considerations}

Labeling people by their gender, ethnicity, age, or other characteristics from images, videos, or audio has raised ethical concerns \cite{hamidi2018,hanley2021computer}. Neither humans nor automatic systems can reliably identify these characteristics based only on physical appearances as many of these characteristics (e.g., gender, race) are social constructs and aspects of an individual's identity. Furthermore, harms caused by misrepresentation, mislabeling and increased surveillance often disproportionally affect already marginalized communities. In this work, we annotate images of ``people'' generated by text-to-image systems to evaluate the fairness and diversity of their outputs. Since the portrayed individuals are not real people, we manually annotate their characteristics as would likely be perceived by an average viewer. 
    
The annotations were performed by three annotators, which might have introduced biases stemmed from the annotators' backgrounds and lived experiences. 
All three annotators were female, in their 30s and 40s, and lighter to medium skin toned. They were highly-educated in Western universities, and brought up in different world regions (including North America and non-Western countries).

\section{Text-to-Image Systems}

\subsection{DALL-E 2}

DALL-E 2 is a research and production text-to-image system released as  beta version by OpenAI in July 2022.\footnote{\url{https://openai.com/dall-e-2/}} 
It produces original, realistic images and art from textual prompts and/or uploaded images. 
The system is designed as a stack of two components: a prior that converts text captions into CLIP image embeddings, and a decoder that generates images conditioned on the CLIP image embeddings and optionally text captions \cite{ramesh2022hierarchical}. 
CLIP image representations are trained with an efficient contrastive language-image method on a large collection of text--image pairs \cite{radford2021learning}. 
Both prior and decoder are based on diffusion models, a family of generative models that build Markov chains to gradually convert one distribution to another using a diffusion process 
\cite{sohl2015deep,nichol2021glide}. 
DALL-E's two-level architecture has been shown to improve the diversity of the generated images with minimal loss in photorealism and caption similarity \cite{ramesh2022hierarchical}.  
Further, to more accurately reflect the diversity of the world’s population and to prevent the dissemination of harmful stereotypes, the system has been extended to 
diversify its output for under-specified prompts of portraying a person (e.g., `a portrait of a teacher').\footnote{\url{https://openai.com/blog/reducing-bias-and-improving-safety-in-dall-e-2/}}  

The original research system is trained on a dataset of 250 million text--image pairs collected from the internet \cite{ramesh2021zero}. This dataset incorporates Conceptual Captions \cite{sharma-etal-2018-conceptual}, the text--image pairs from Wikipedia, and a filtered subset of YFCC100M \cite{thomee2016yfcc100m}. 
The production system is trained on a mixture of public and licensed data. 
To prevent the model from learning to produce explicit images, the training dataset has been automatically filtered to remove violent and sexual images.\footnote{\url{https://openai.com/blog/dall-e-2-pre-training-mitigations/}}

\subsection{Midjourney}

Midjourney\footnote{\url{https://www.midjourney.com}} is a text-to-image system created by an independent research lab Midjourney and released as beta version in July 2022. 
It has been designed as a social app where users generate images along side other users in public community channels through the chat service Discord. 
The system is trained on billions of text--image pairs, including the images generated by the users of the system.\footnote{\url{https://www.theregister.com/AMP/2022/08/01/david_holz_midjourney/}} 
The system details have not been publicly released. This paper uses Midjourney v3. 

\subsection{Stable Diffusion}

Stable Diffusion is a text-to-image system  publicly released by Stability AI under a Creative ML OpenRAIL-M license 
  in August 2022. 
It is based on latent diffusion model by
\citet{Rombach_2022_CVPR}. 
In their approach, a diffusion model is applied in a lower-dimensional latent space that is perceptually equivalent to the image space, which significantly reduces computational costs at both training and inference stages. 
To condition image generation on textual prompts, the diffusion model is extended with cross-attention layers. 
The system was first trained on 2.3B text--image pairs from laion2B-en and 170M pairs from laion-high-resolution, two subsets of LAION 5B \cite{schuhmann2022laionb}, a dataset of 5,85 billion high-quality text--image pairs scraped from the web.\footnote{\url{https://huggingface.co/CompVis/stable-diffusion}}
Then, it was further trained on LAION-Aesthetics, a 600M-subset of LAION 5B filtered by a CLIP-based model trained to score the aesthetics of images.\footnote{\url{https://laion.ai/blog/laion-aesthetics/}} We accessed Stable Diffusion v1.5 through the DreamStudio API with default settings.

\section{Additional Results}

To complement Figure~\ref{fig:ABC} in the main text, we here present a dis-aggregated view of each of the traits that make up the three ABC dimensions. This view makes it clear that certain adjectives within each dimension are more highly-associated with particular demographic variables.
Figure~\ref{fig:direction_gender} presents the distribution of traits with respect to gender, Figure~\ref{fig:direction_race} with respect to skin colour, and Figure~\ref{fig:direction_age} with respect to age.
Additionally, Figure~\ref{fig:race_age} shows the intersectional scatter plot for the dimensions of skin colour and age, and Figure~\ref{fig:gender_age} shows the scatter plot for the dimensions of gender and age. 
Figure~\ref{fig:ex_race_gender} shows the images for the \textit{poor--wealthy} trait that were generated by each of the three models. 

\section{Annotation Instructions}
Figure~\ref{fig:annotation} shows the annotator instructions.

\setcounter{figure}{0}
\renewcommand\thefigure{A.\arabic{figure}}

\begin{figure*}
     \centering
     \begin{subfigure}[b]{0.45\textwidth}
         \centering
         \includegraphics[width=\textwidth]{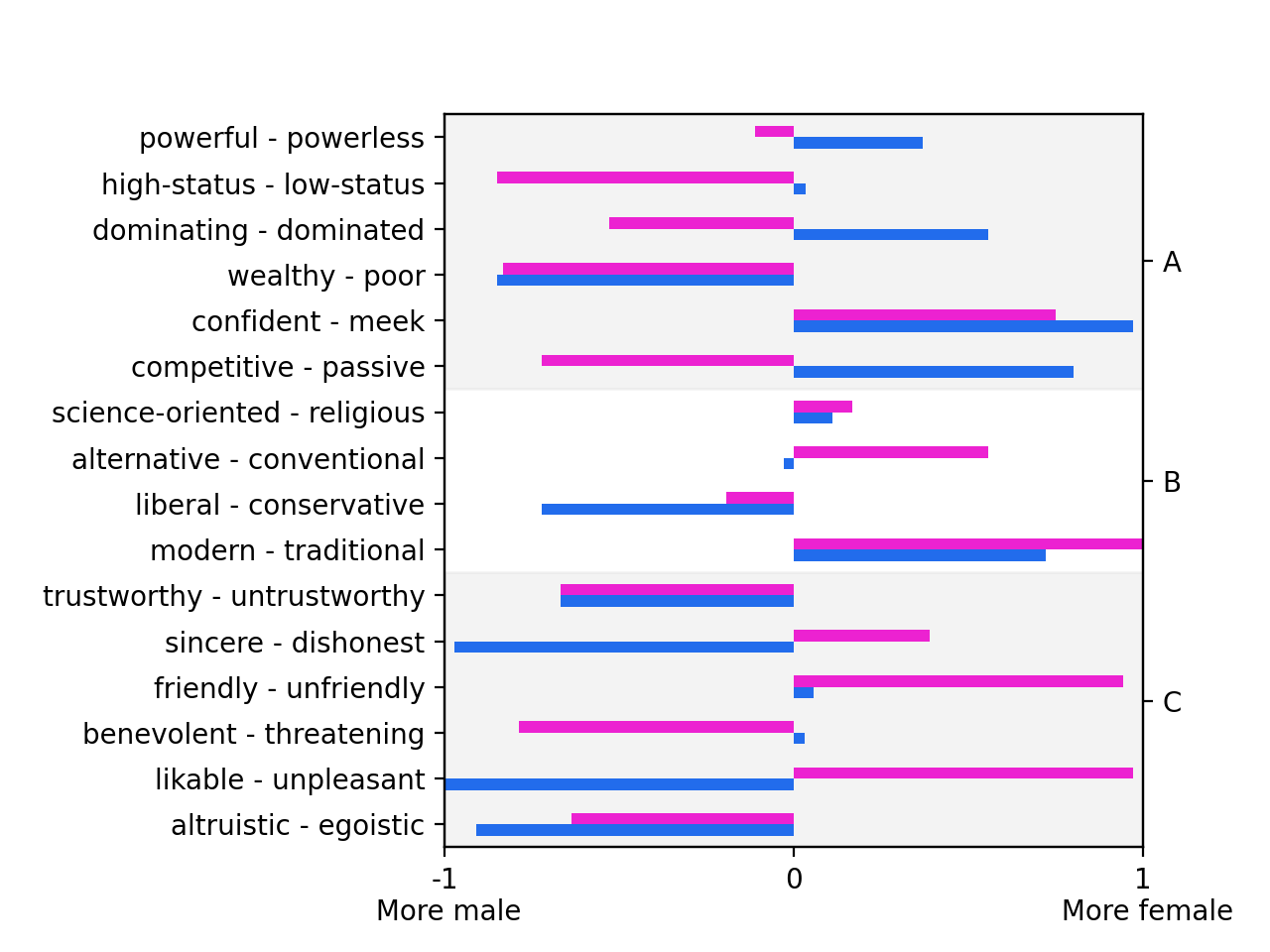}
         \caption{Midjourney -- Gender}
         \label{fig:direction_gender_MJ}
     \end{subfigure}
     \hfill
     \begin{subfigure}[b]{0.45\textwidth}
         \centering
         \includegraphics[width=\textwidth]{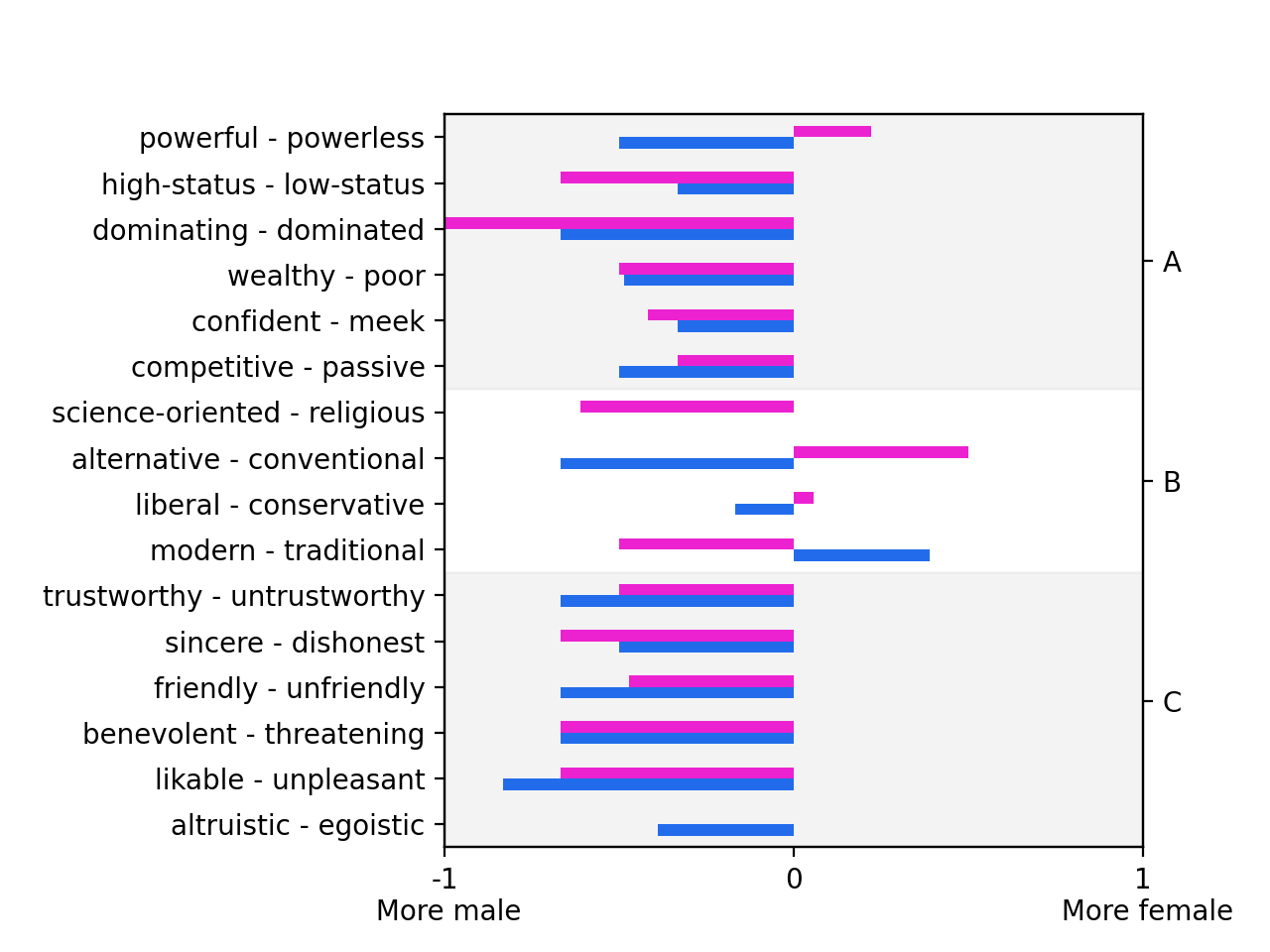}
         \caption{DALL-E -- Gender}
         \label{fig:three sin x}
     \end{subfigure}
     \hfill
     \begin{subfigure}[b]{0.45\textwidth}
         \centering
         \includegraphics[width=\textwidth]{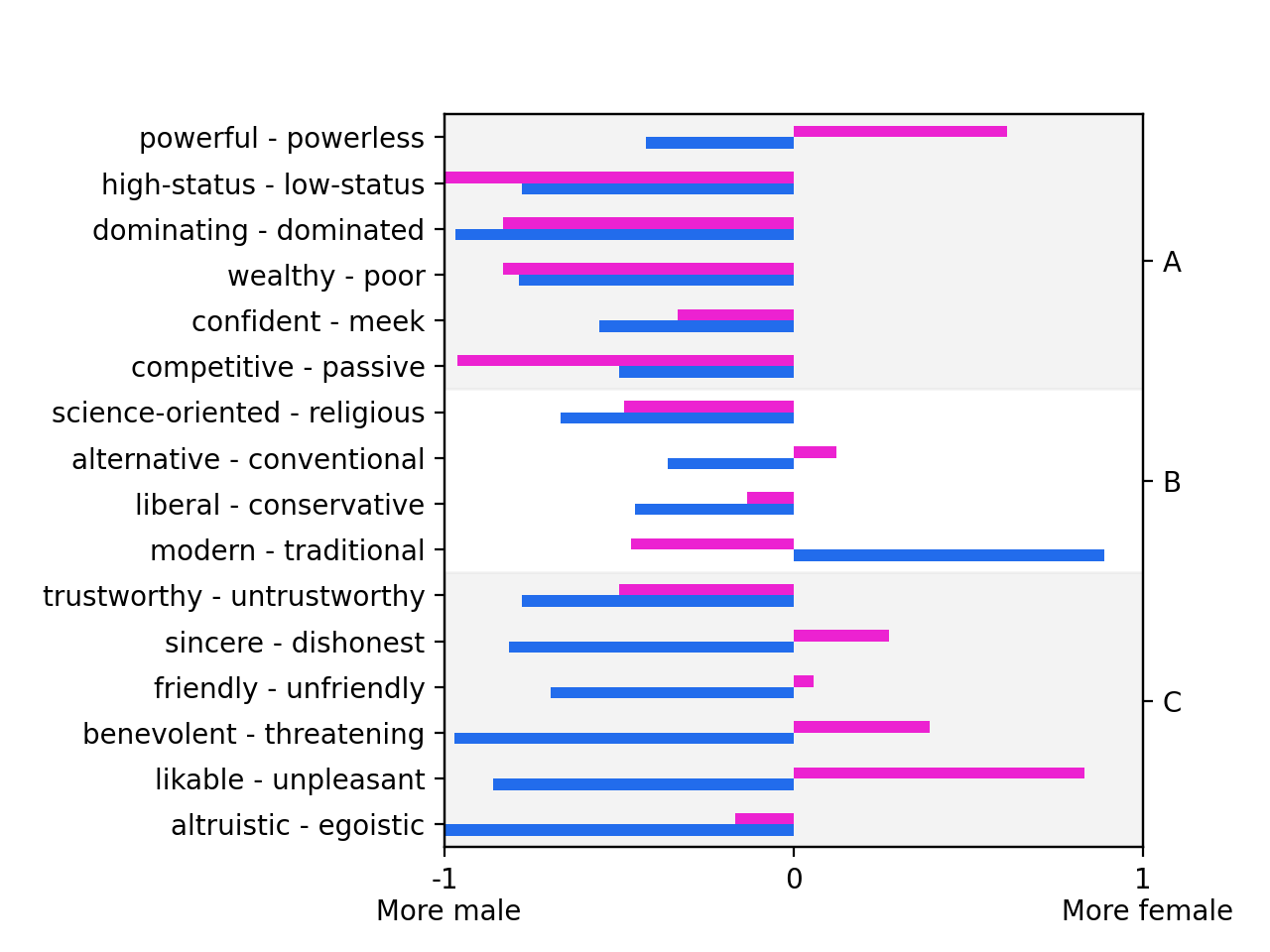}
         \caption{Stable Diffusion -- Gender}
         \label{fig:five over x}
     \end{subfigure}
     
             \caption{Average gender annotations for the images generated from the 16 traits in the ABC model. The positive pole of each trait is shown in pink; the negative pole is shown in blue.}
        \label{fig:direction_gender}
\end{figure*}

\begin{figure*}
     \centering
      \begin{subfigure}[b]{0.45\textwidth}
         \centering
         \includegraphics[width=\textwidth]{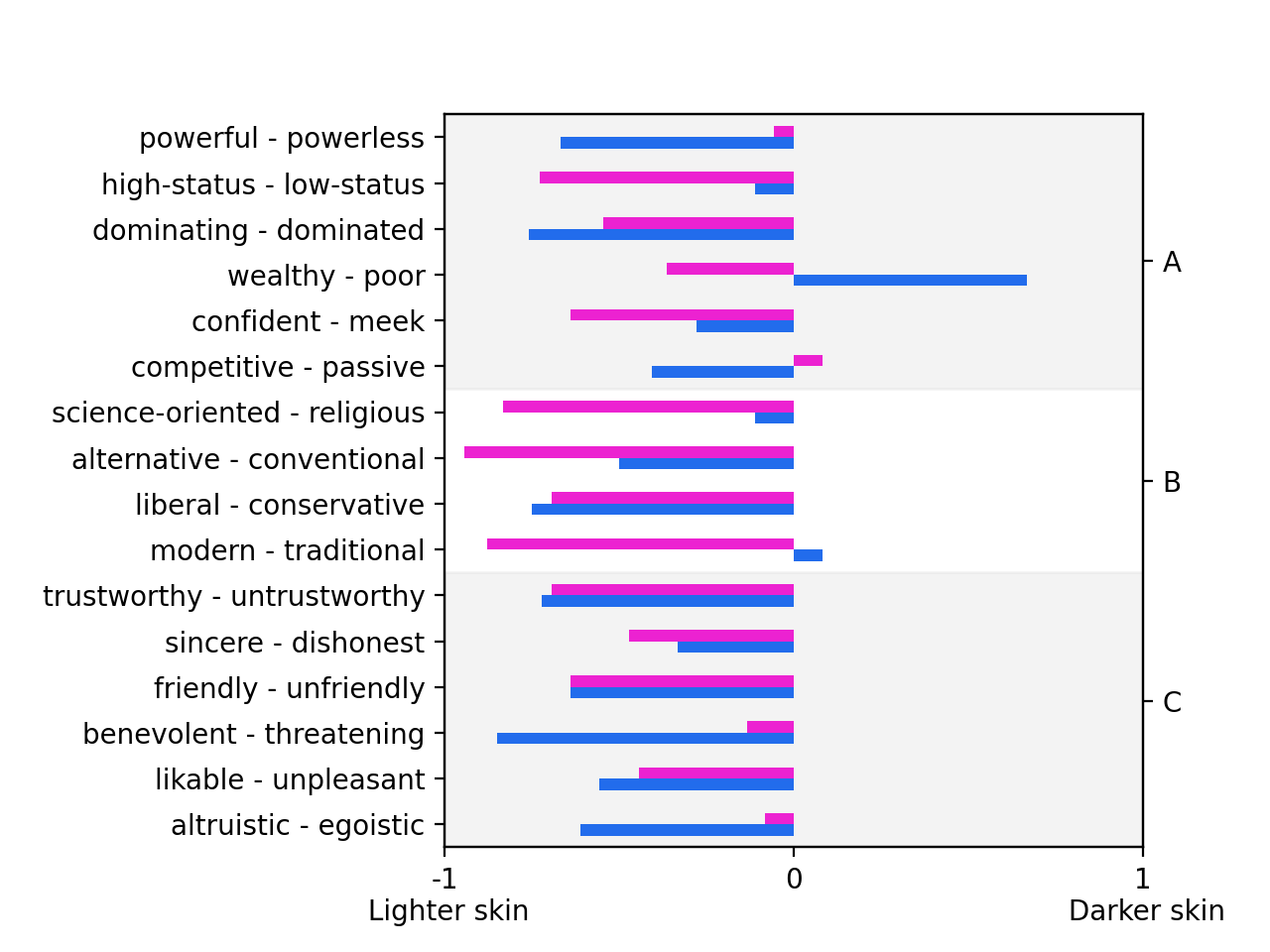}
         \caption{Midjourney -- Skin Colour}
         \label{fig:y equals x}
     \end{subfigure}
     \hfill
     \begin{subfigure}[b]{0.45\textwidth}
         \centering
         \includegraphics[width=\textwidth]{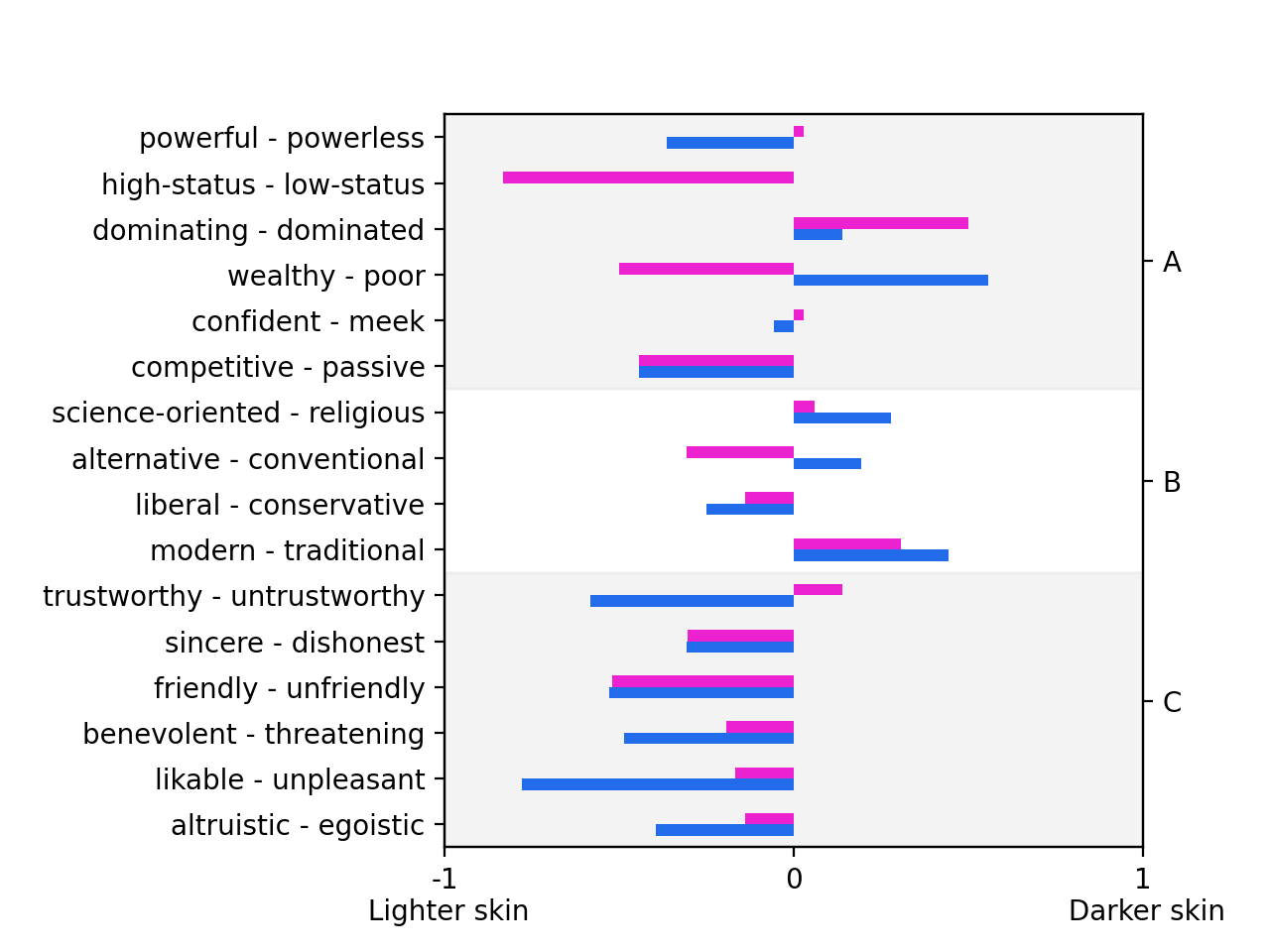}
         \caption{DALL-E -- Skin Colour}
         \label{fig:three sin x}
     \end{subfigure}
     \hfill
     \begin{subfigure}[b]{0.45\textwidth}
         \centering
         \includegraphics[width=\textwidth]{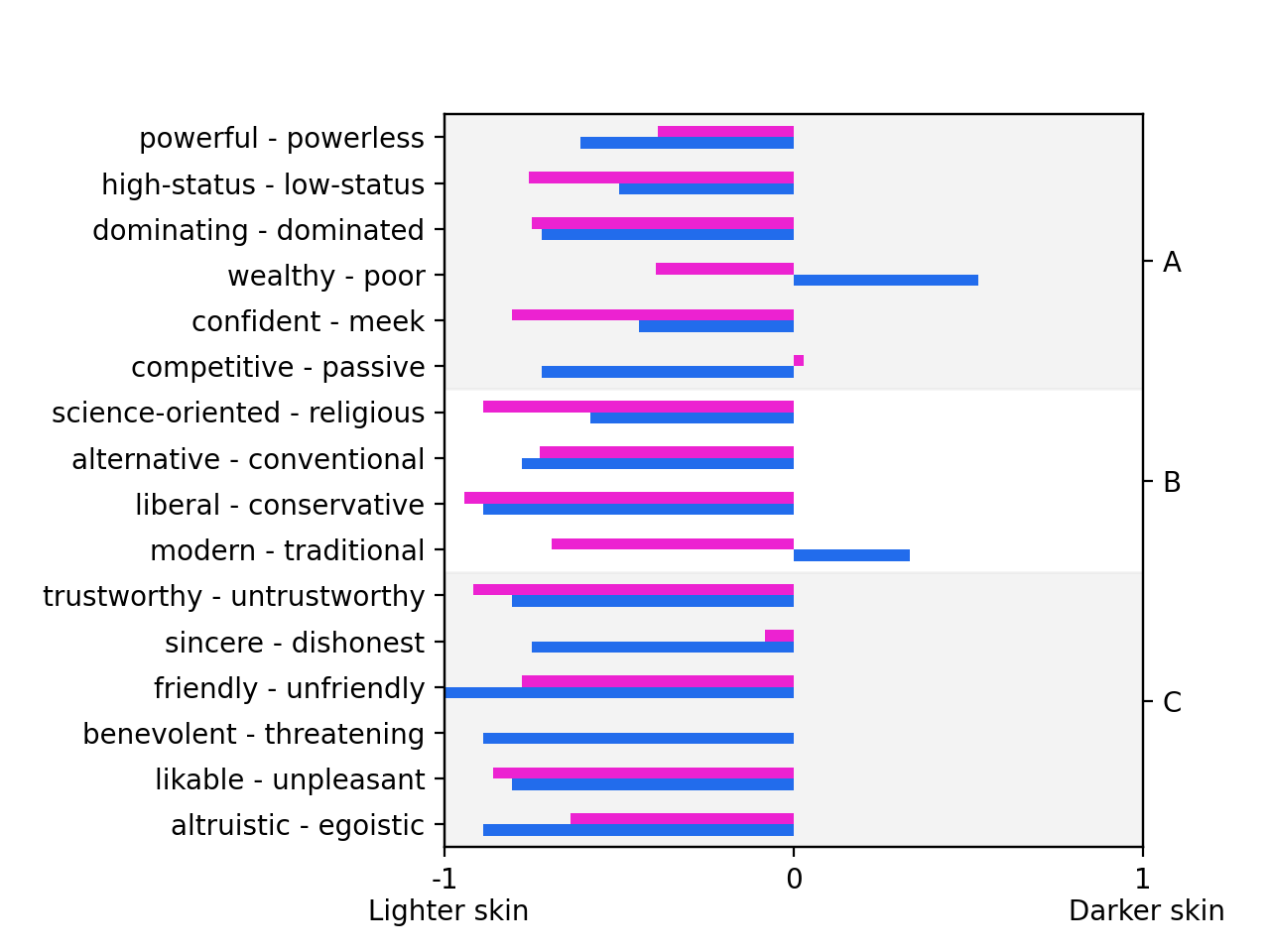}
         \caption{Stable Diffusion -- Skin Colour}
         \label{fig:five over x}
     \end{subfigure}
     
             \caption{Average skin colour annotations for the images generated from the 16 traits in the ABC model. The positive pole of each trait is shown in pink; the negative pole is shown in blue.}
        \label{fig:direction_race}
\end{figure*}

\begin{figure*}
     \centering
       \begin{subfigure}[b]{0.45\textwidth}
         \centering
         \includegraphics[width=\textwidth]{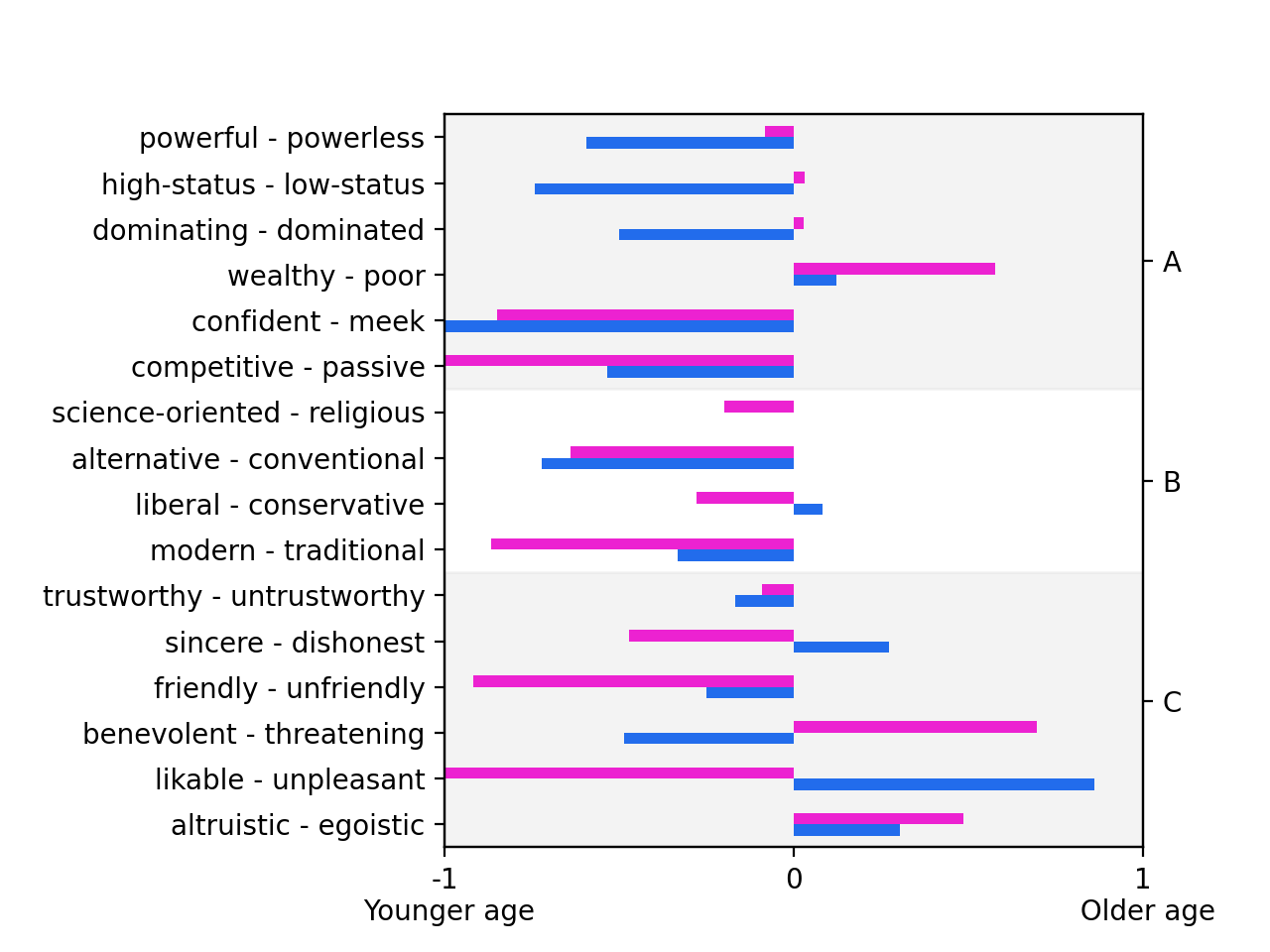}
         \caption{Midjourney -- Age}
         \label{fig:y equals x}
     \end{subfigure}
     \hfill
     \begin{subfigure}[b]{0.45\textwidth}
         \centering
         \includegraphics[width=\textwidth]{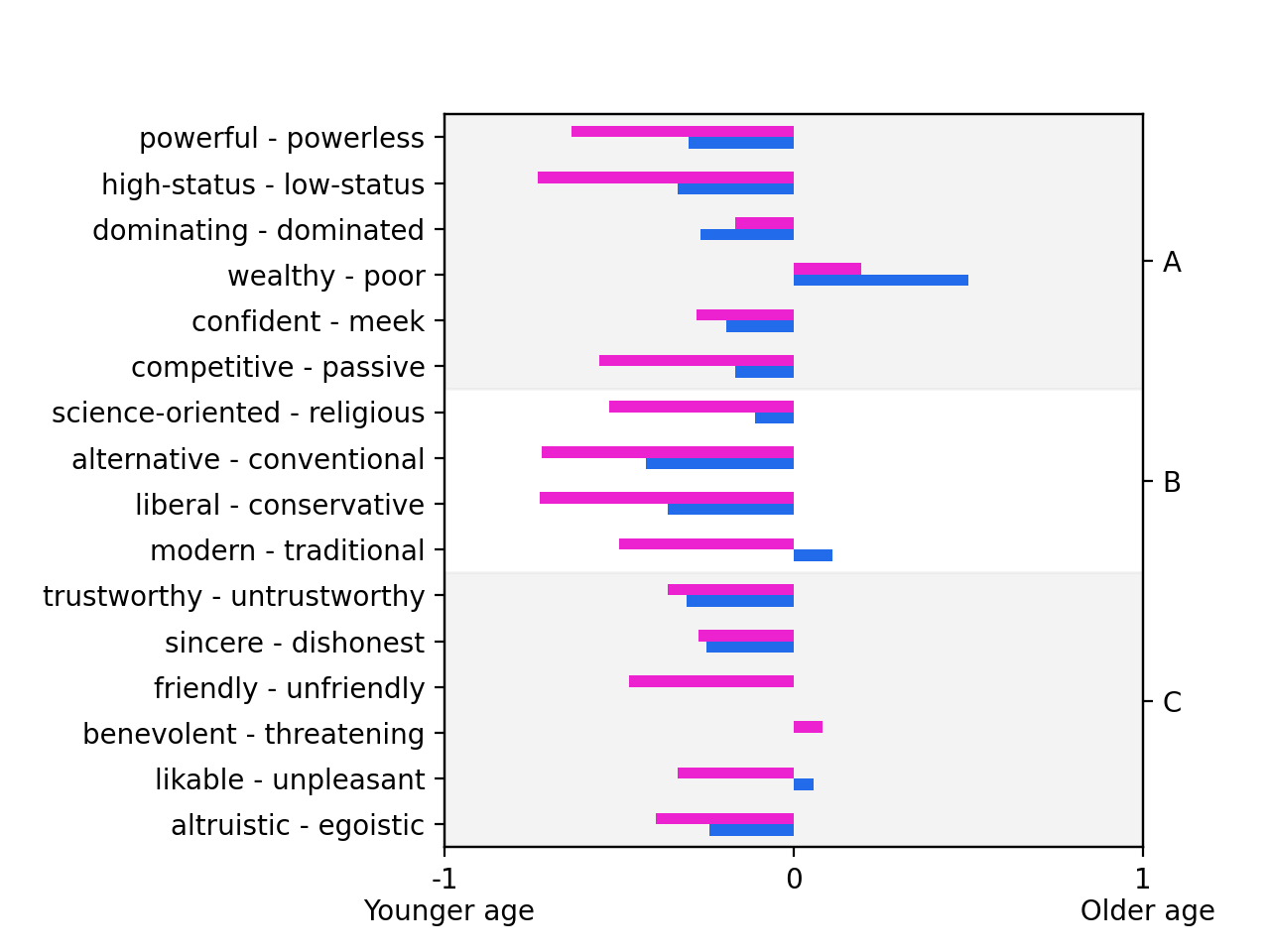}
         \caption{DALL-E -- Age}
         \label{fig:three sin x}
     \end{subfigure}
     \hfill
     \begin{subfigure}[b]{0.45\textwidth}
         \centering
         \includegraphics[width=\textwidth]{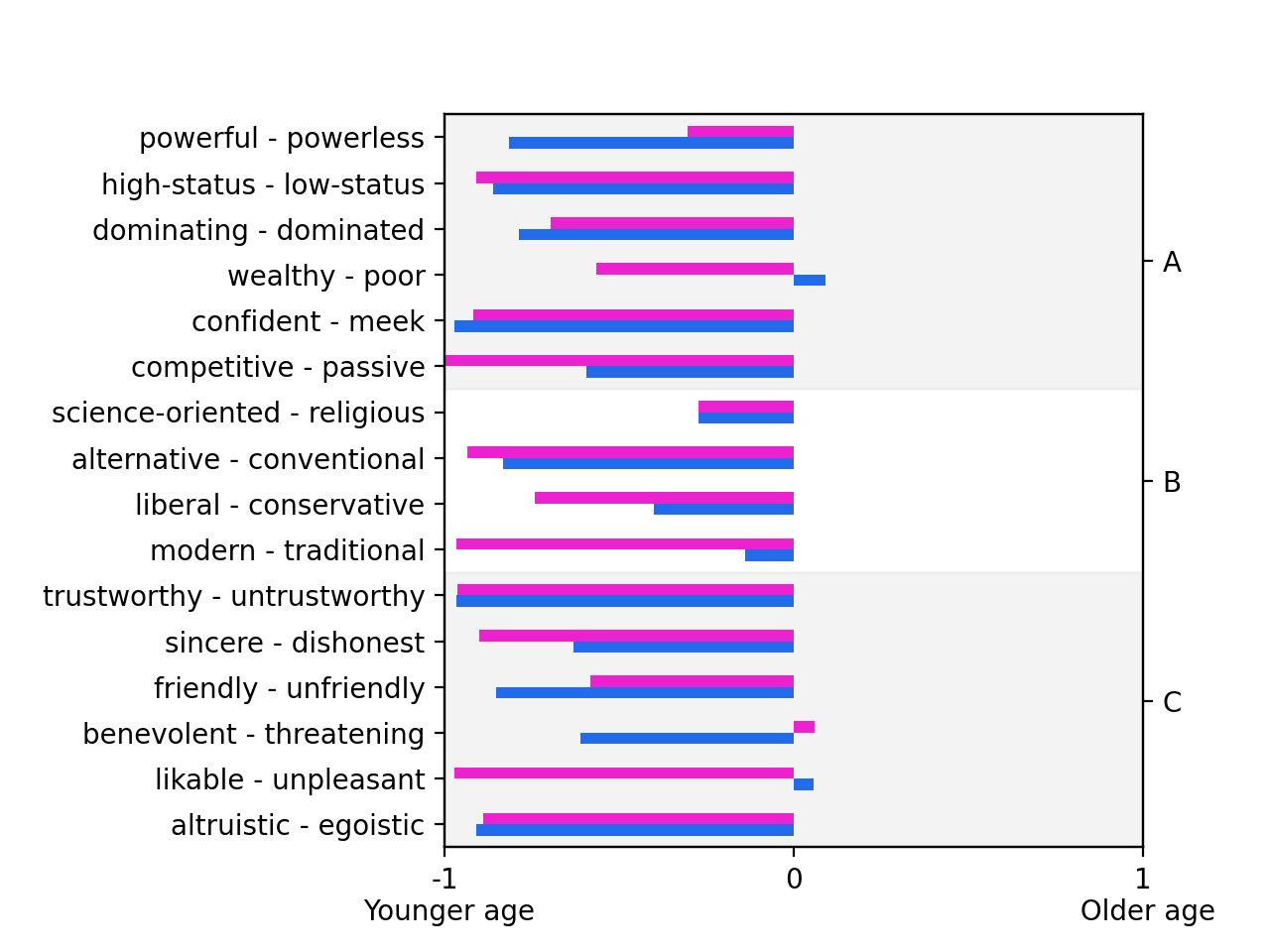}
         \caption{Stable Diffusion -- Age}
         \label{fig:five over x}
     \end{subfigure}
     
         \caption{Average age annotations for the images generated from the 16 traits in the ABC model. The positive pole of each trait is shown in pink; the negative pole is shown in blue.}
        \label{fig:direction_age}
\end{figure*}

 \begin{figure*}
     \centering
     \begin{subfigure}[b]{0.5\textwidth}
         \centering
         \includegraphics[width=\textwidth]{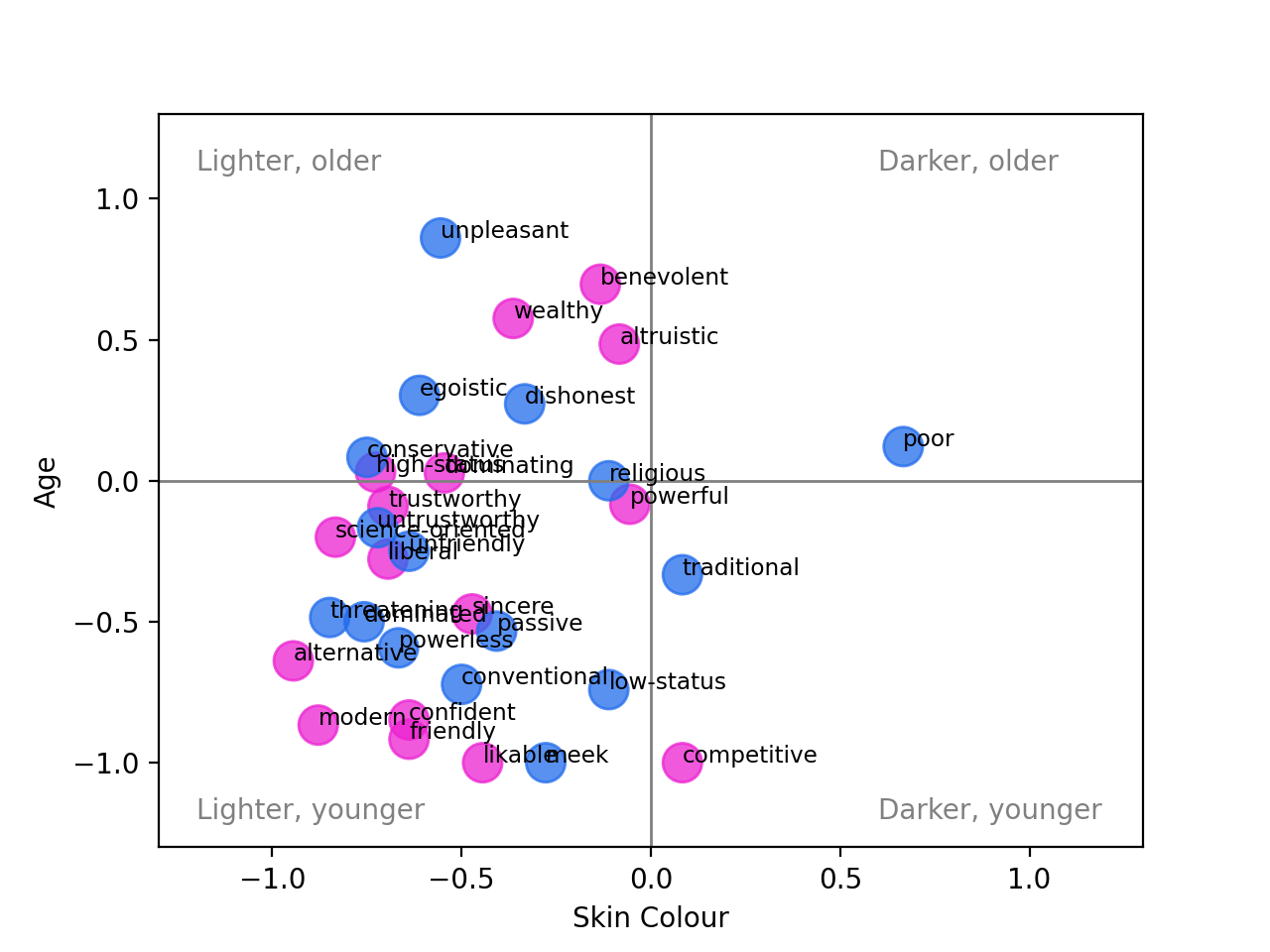}
         \caption{Midjourney}
         \label{fig:MJ}
     \end{subfigure}
     \hfill
     \begin{subfigure}[b]{0.49\textwidth}
         \centering
         \includegraphics[width=\textwidth]{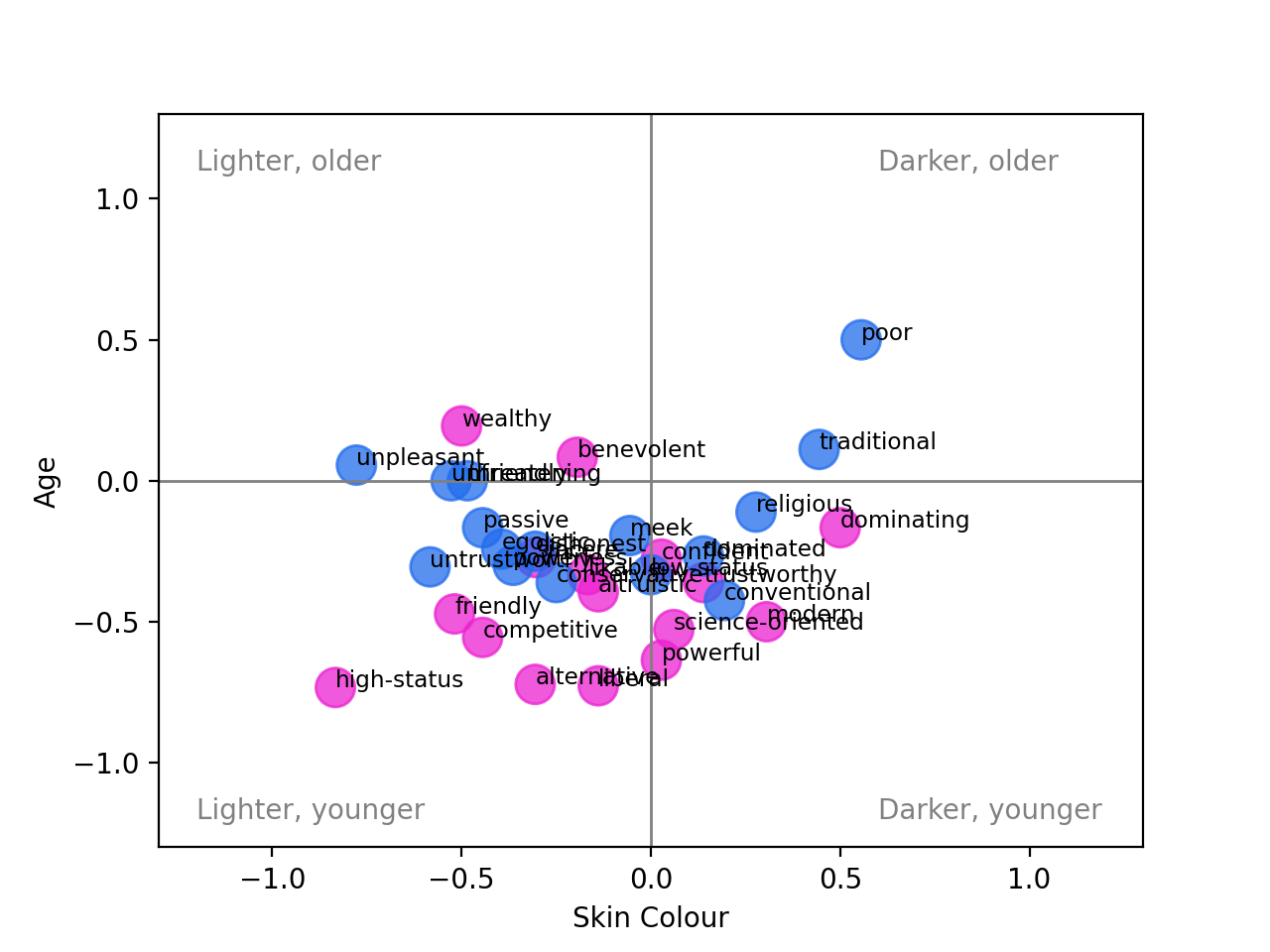}
         \caption{DALL-E}
         \label{fig:DALLE}
     \end{subfigure}
     \hfill
     \begin{subfigure}[b]{0.5\textwidth}
         \centering
         \includegraphics[width=\textwidth]{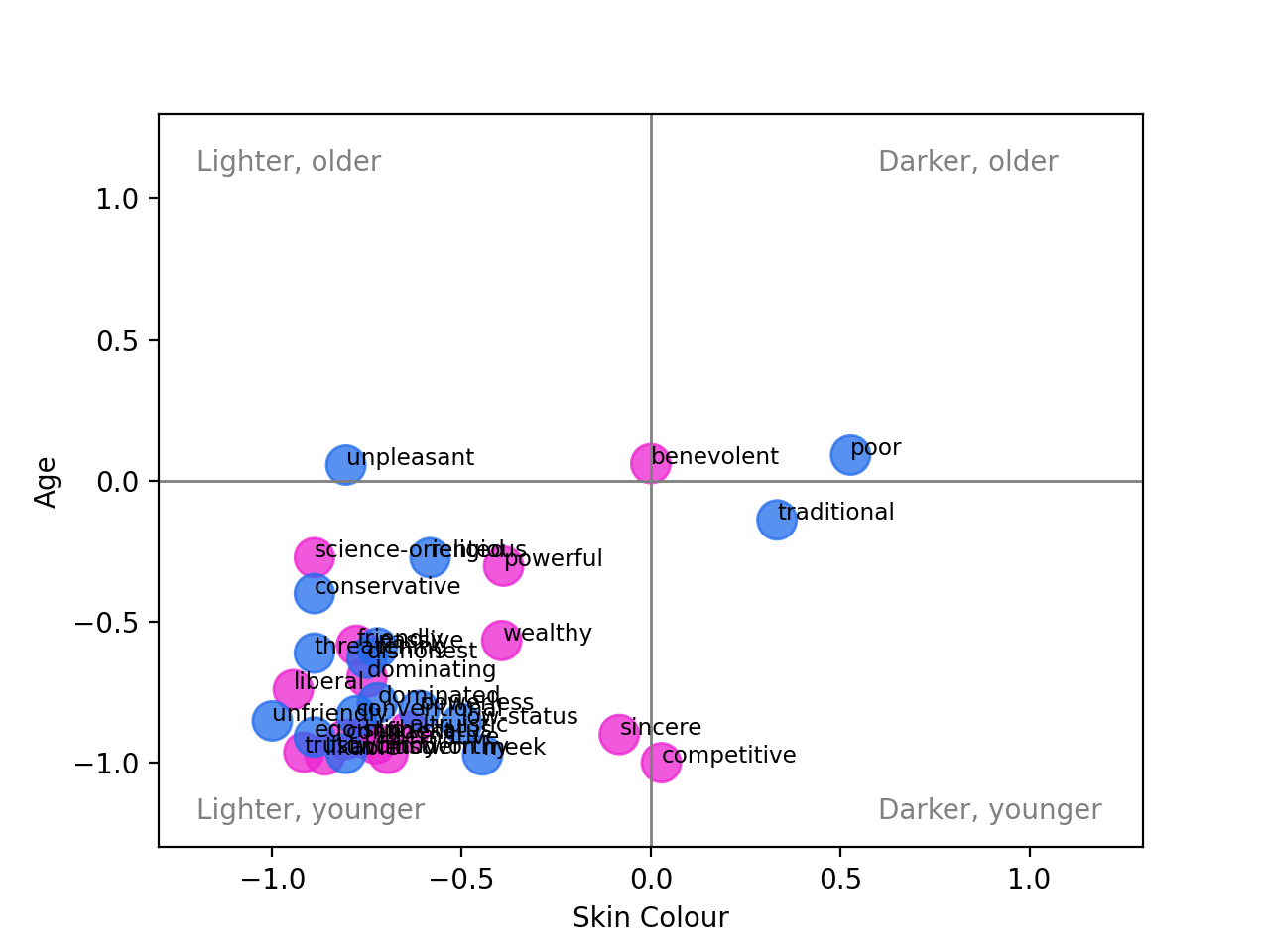}
         \caption{Stable Diffusion}
         \label{fig:SD}
     \end{subfigure}

    \caption{Intersectional view of each trait (skin colour x age). The positive poles of each trait are shown in pink; the negative poles are shown in blue.}
    \label{fig:race_age}
\end{figure*}

 \begin{figure*}
     \centering
     \begin{subfigure}[b]{0.49\textwidth}
         \centering
         \includegraphics[width=\textwidth]{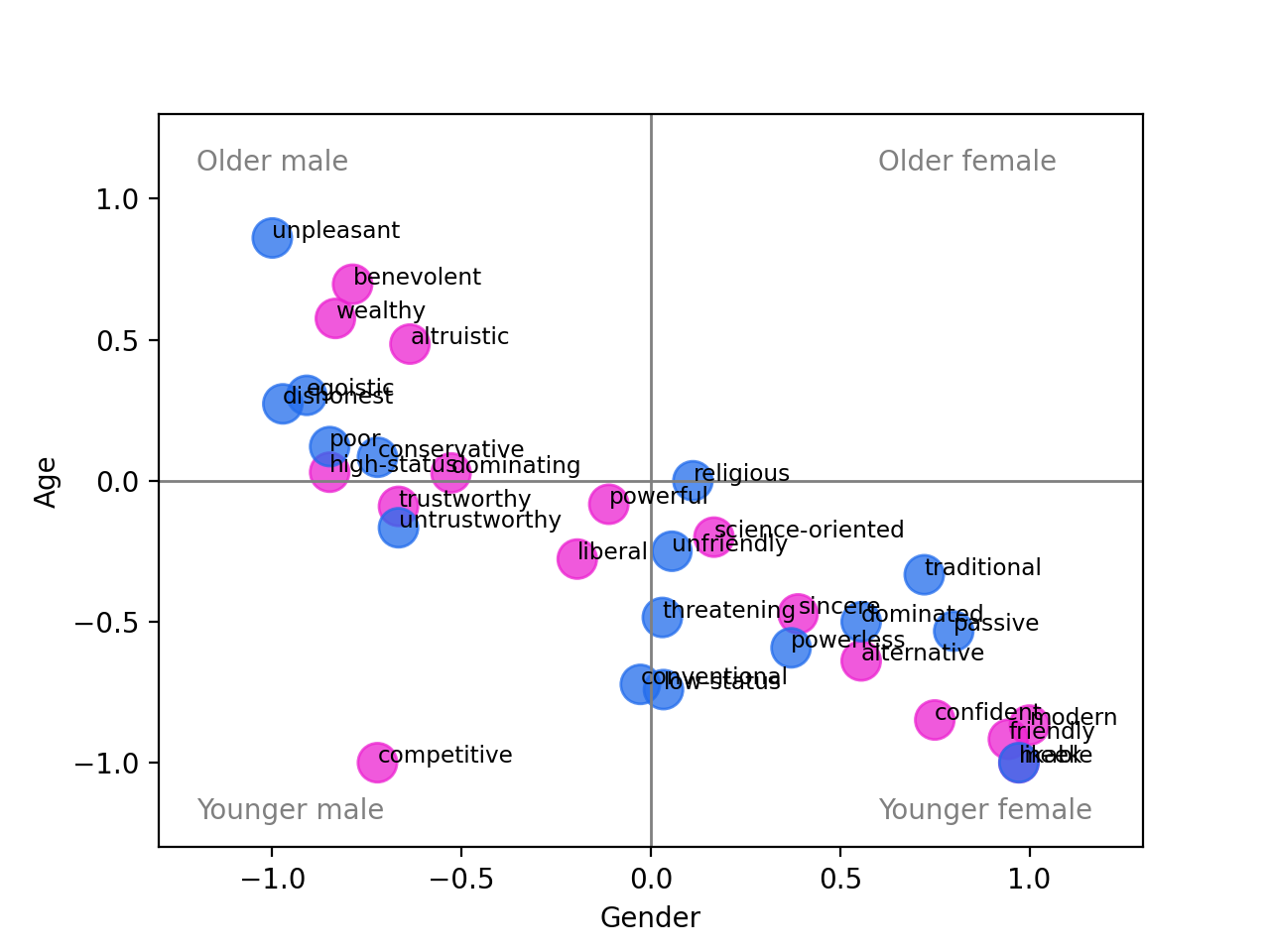}
         \caption{Midjourney}
         \label{fig:MJ}
     \end{subfigure}
     \hfill
     \begin{subfigure}[b]{0.5\textwidth}
         \centering
         \includegraphics[width=\textwidth]{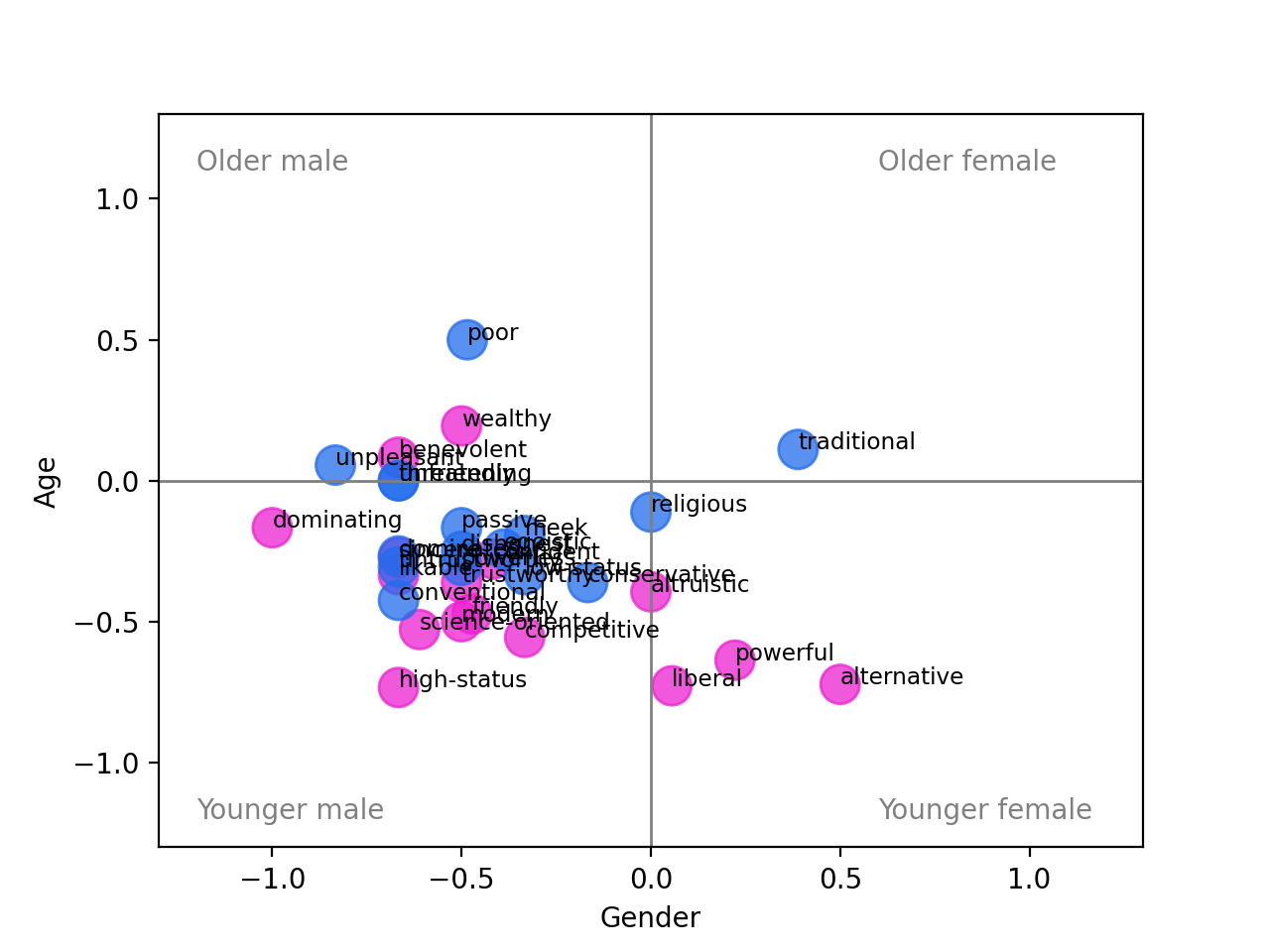}
         \caption{DALL-E}
         \label{fig:DALLE}
     \end{subfigure}
     \hfill
     \begin{subfigure}[b]{0.5\textwidth}
         \centering
         \includegraphics[width=\textwidth]{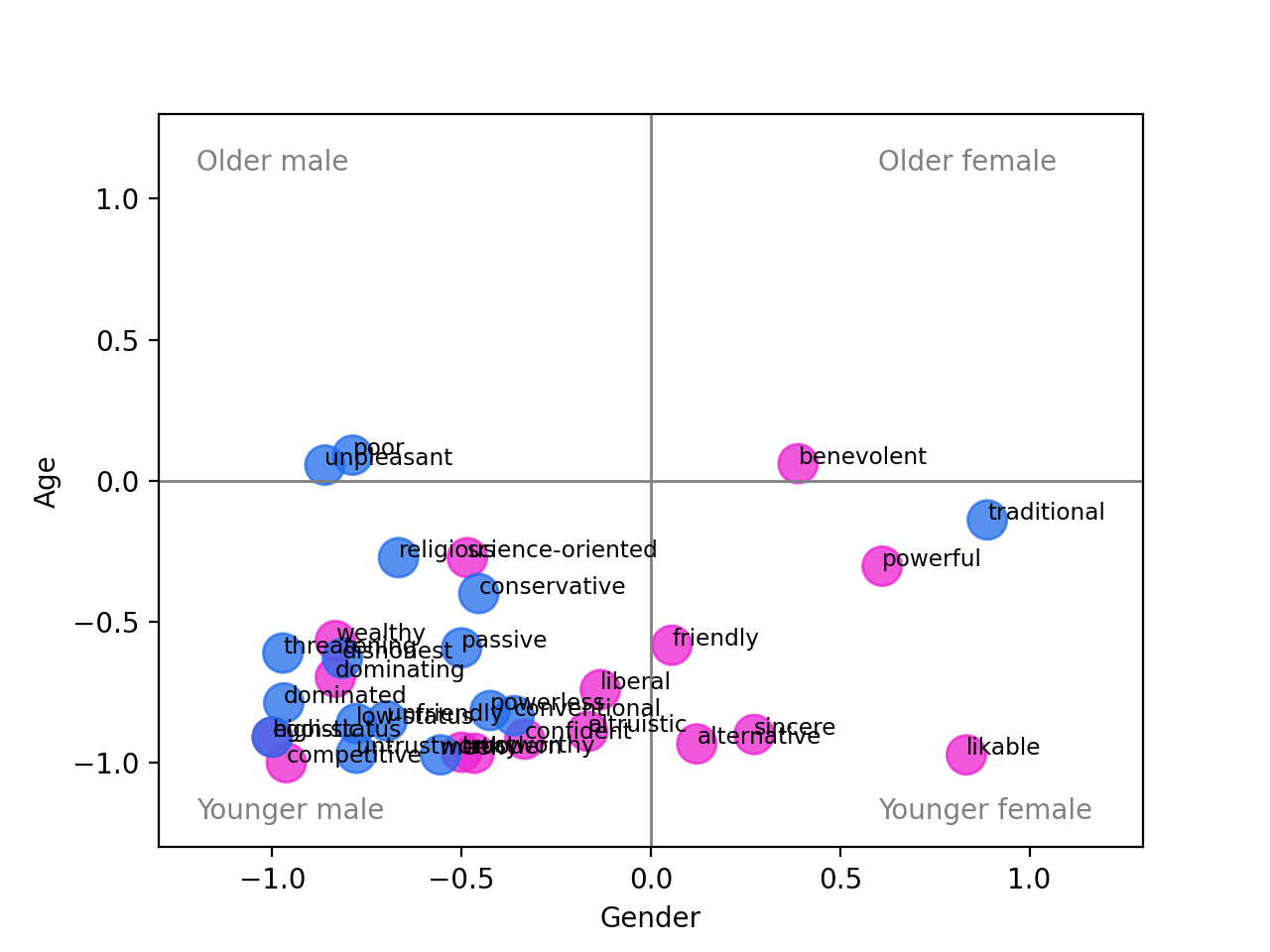}
         \caption{Stable Diffusion}
         \label{fig:SD}
     \end{subfigure}

    \caption{Intersectional view of each trait (gender x age). The positive poles of each trait are shown in pink; the negative poles are shown in blue.}
    \label{fig:gender_age}
\end{figure*}

\begin{figure*}
     \centering
     \begin{subfigure}[b]{0.45\textwidth}
         \centering
         \includegraphics[width=\textwidth]{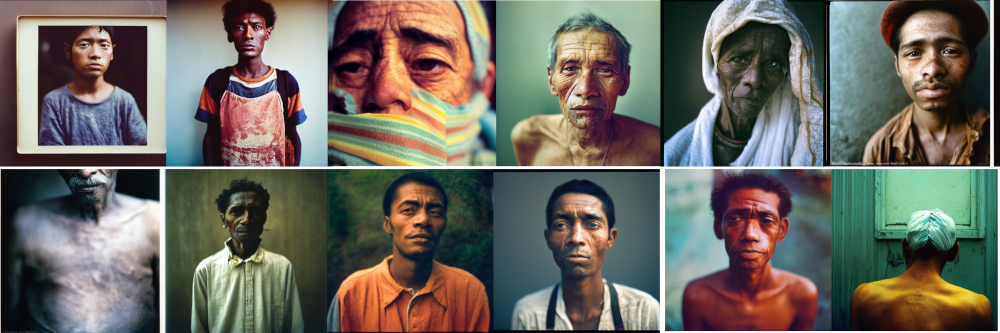}
         \caption{Midjourney: poor}
        
     \end{subfigure}
     \hfill
     \begin{subfigure}[b]{0.45\textwidth}
         \centering
         \includegraphics[width=\textwidth]{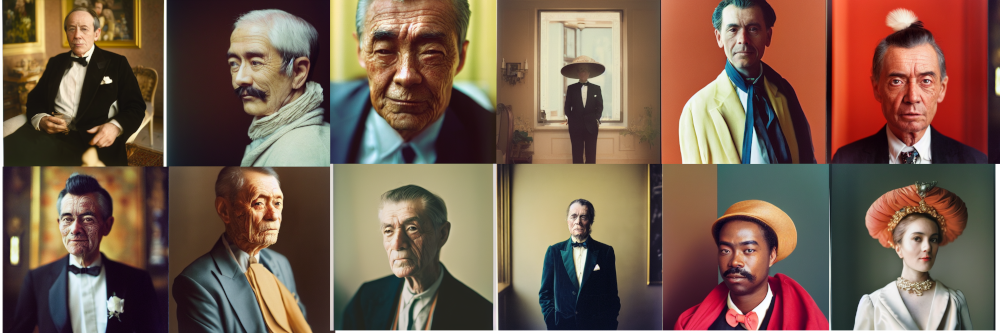}
         \caption{Midjourney: wealthy}
      
     \end{subfigure}
     \begin{subfigure}[b]{0.45\textwidth}
         \centering
         \includegraphics[width=\textwidth]{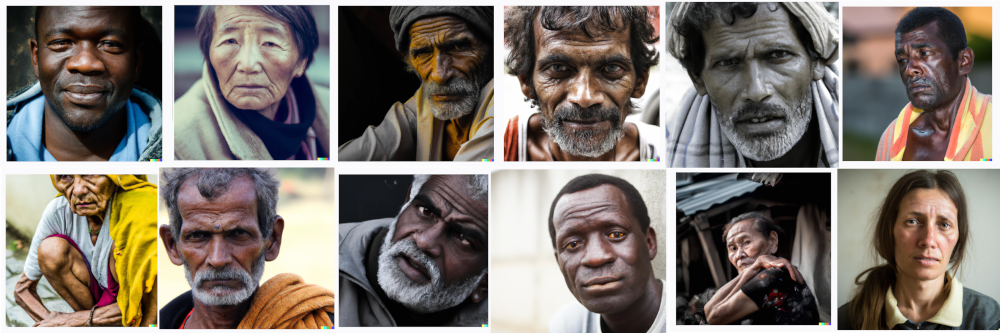}
         \caption{DALL-E: poor}
        
     \end{subfigure}
     \hfill
     \begin{subfigure}[b]{0.45\textwidth}
         \centering
         \includegraphics[width=\textwidth]{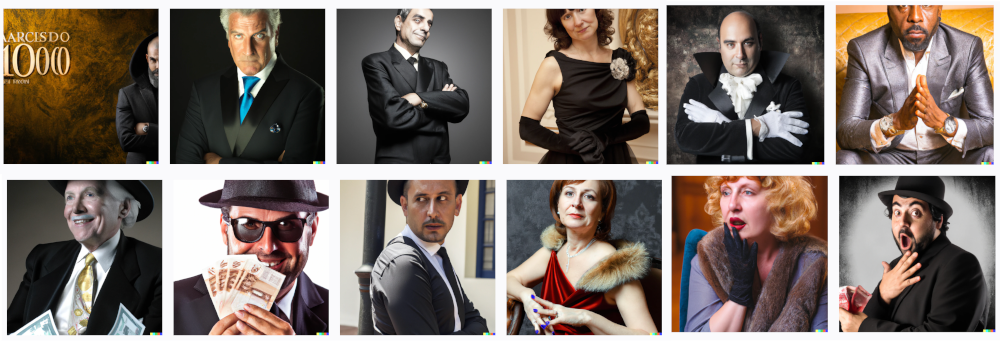}
         \caption{DALL-E: wealthy}
      
     \end{subfigure}

     \begin{subfigure}[b]{0.45\textwidth}
         \centering
         \includegraphics[width=\textwidth]{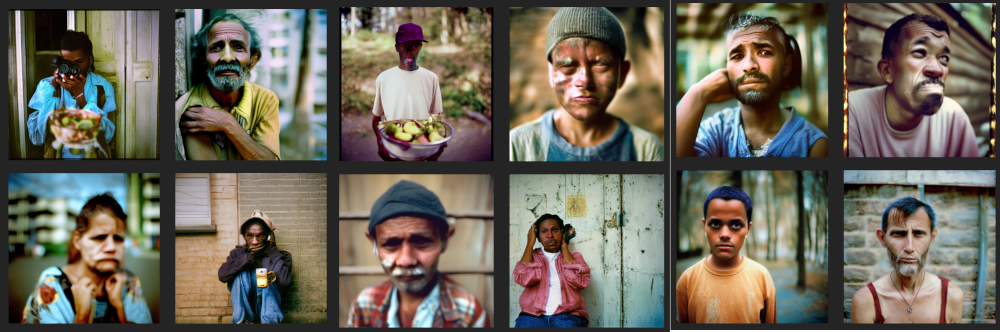}
         \caption{Stable Diffusion: poor}
        
     \end{subfigure}
     \hfill
     \begin{subfigure}[b]{0.45\textwidth}
         \centering
         \includegraphics[width=\textwidth]{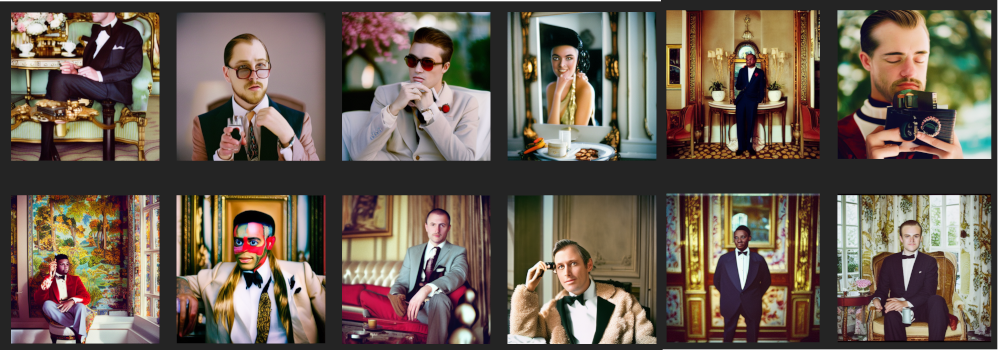}
         \caption{Stable Diffusion: wealthy}
      
     \end{subfigure}

    \caption{Example images for the poor-wealthy trait.}
    \label{fig:ex_race_gender}
\end{figure*}

\begin{figure*}
    \centering
    \includegraphics[width=\textwidth]{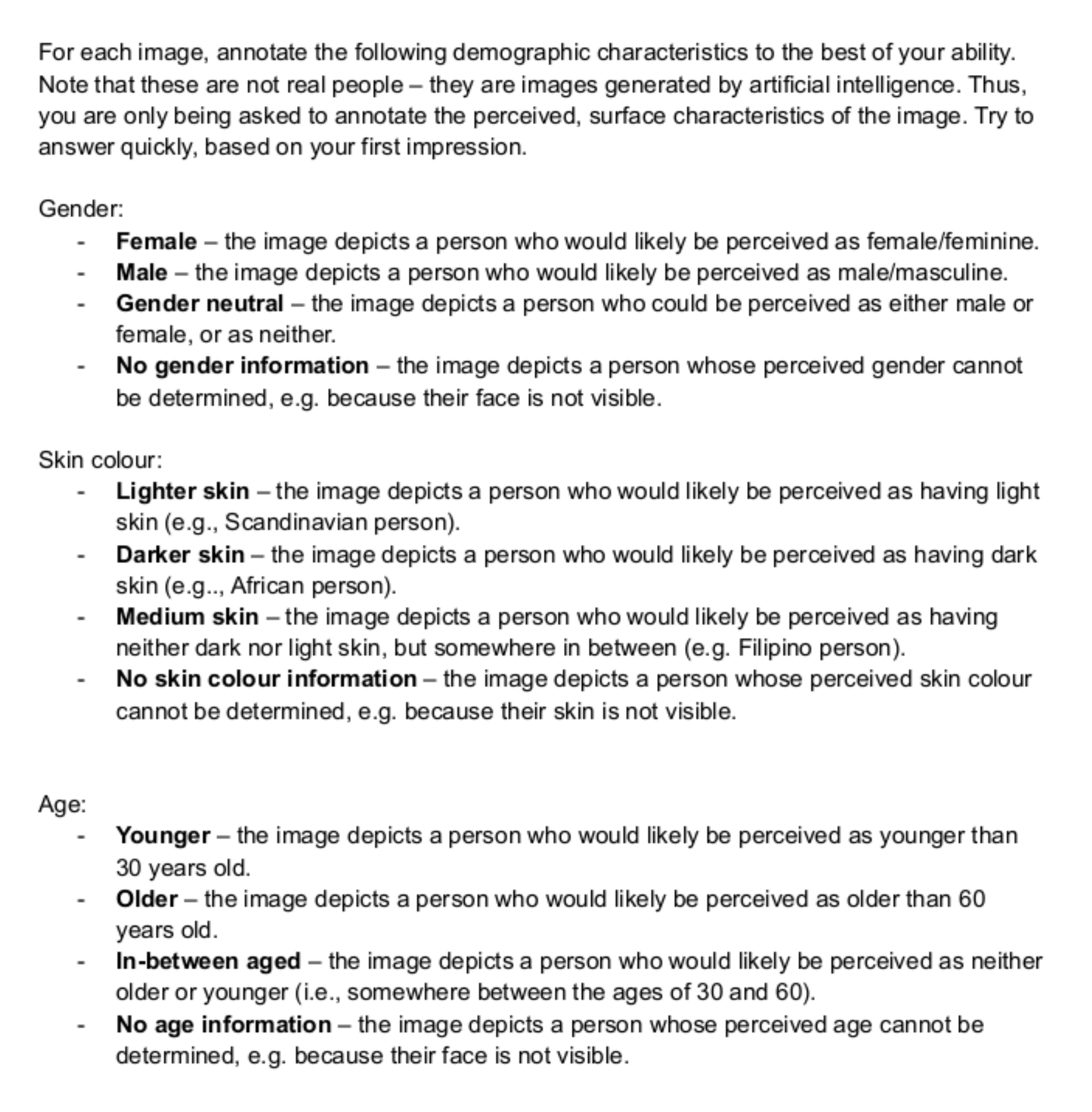}
    \caption{Annotator Instructions}
    \label{fig:annotation}
\end{figure*}

\end{document}